\documentclass[showkeys,prc,twocolumn]{revtex4}
\usepackage{epsfig}
\usepackage{graphicx}
\usepackage{amssymb}
\usepackage{amsmath}
\usepackage{hyperref}
\usepackage{natbib}

\usepackage{color}

\begin{document}

\title{Coherent Neutron Scattering in Polycrystalline Deuterium and its Implications for Ultracold Neutron Production}
\date{\today}

\author{C.-Y. Liu$^1$}
\email{CL21@indiana.edu}
\author{A.R. Young$^2$}
\author{C. Lavelle$^1$}
\author{D. Salvat$^1$}
\affiliation{1. Physics Department, Indiana University, Bloomington, IN 47408}
\affiliation{2. Physics Department, North Carolina State University, Raleigh, NC 27695}

\begin{abstract}
This paper presents a calculation of the neutron 
cross-sections in solid materials (used in practical neutron sources) with a large coherent scattering contribution. 
In particular, the dynamic structure function S(Q, $\omega$) of polycrystalline ortho-D$_2$ is evaluated using a Monte-Carlo calculation that performs an average over scattering angles relative to crystal axes in random orientations. 
This method uses an analytical dispersion function with force constants derived from neutron scattering data of single crystal D$_2$ in the framework of an axially symmetric force tensor. 
The resulting two dimensional map of S(Q, $\omega$) captures details of the phonon branches as well as the molecular rotations, that can be compared directly to data from inelastic neutron scattering on polycrystalline D$_2$.  
This high resolution information is used to calculate the absolute cross-sections of production and upscattering loss of ultracold neutron (UCN). The resulting scattering cross-sections are significantly different, especially for UCN upscattering, from the previous predictions using the approach centered on the incoherent approximation.  
\end{abstract}



\maketitle

\section{\label{sec:intro}Introduction}
Neutron moderation is a well established technique, developed 
for applications in fission reactors\cite{Glasstone52} and nuclear weapons. Interest in this topic has been revived with the development of new neutron scattering facilities for investigation of materials, new sources of ultracold neutrons for fundamental physics, as well as new deep-underground laboratories which require extensive shielding. 
In the latter case, neutrons generated in the surrounding rock via the spallation process initiated by high energy cosmic rays are a major source of background.
On the pure academic front, neutron moderation can be put in a broader context of phase space compression, 
developed for several different types of particle beams in various areas of physics, from stochastic electron cooling in high energy particle accelerators on Tera-eV scales to laser cooling of neutral atoms to form Bose-Einstein condensates on pico-eV scales.
The phase space compression of neutrons is especially challenging. 
Because the electromagnetic interaction of the neutron is relatively weak due to its zero charge and small magnetic moment, the usual mechanisms of phase space compression for charged particles do not apply. Instead, the nuclear force couples neutrons to individual nuclei in the interacting medium. Upon scattering, the nuclear recoil leads to rapid moderation for fast neutrons, but stops to be effective for neutrons with energy below a few tens of milli-eV. For neutrons of particular interests to fundamental research, such as cold, very-cold, and ultra-cold neutrons (with energies ranging from milli-eV to nano-eV), the only efficient way to date to increase neutron phase space density is through inelastic collisions in a cold medium, so-called neutron moderator. 

To study the process of neutron cooling and estimate the neutron flux for the broad range of applications mentioned above, standard simulation codes are available including the widely-used Monte Carlo N-Particle Transport Code (MCNP)\cite{mcnp5}. The code was developed in the 1960s at Los Alamos National Laboratory, and has since been extensively bench-marked with experimental data; the physics models and scattering kernels used have been refined with every new release. 
In MCNP, moderations of neutrons in a medium are modeled by the Boltzmann equation that details the evolution of the phase space distribution function of neutrons. The Boltzmann equation includes neutron diffusion, neutron absorption loss, and inelastic processes in which neutrons downscatter with energy loss or upscatter with energy boost. In this formalism, the scattering kernel determines the probability for inelastic processes. For fast neutrons (with energies from eV to MeV), nuclear resonances strongly influence the scattering and absorption processes. Away from the resonances, the neutron energy loss through nuclear recoil is calculated with a simple free gas model, in which the time scale of collisions is so short that nuclei are treated as free. 
For neutrons with wavelength larger than the separation between scattering nuclei, however, thermal neutron scattering kernels that include coherent diffraction and collective excitations in bulk materials have only been incorporated for limited materials~\cite{MacFarlane94}.     
Whereas nuclear cross-sections for fast neutron reactions are detailed for almost every stable isotope, there exist only a handful of materials with extensive treatments that include condensed matter effects beyond the simple nuclear recoil in the free gas model. 


Hydrogen and deuterium in liquid and solid forms are of particular interest, as they are the materials of choice to construct cold neutron moderators used in research reactors and spallation neutron sources. 
The large nuclear cross-section, together with the maximum energy loss in recoil kinematics of n-$^1$H scattering, leads to rapid cool-down of incident neutrons. However, the $^1$H(n,$\gamma$)$^2$H reaction results in significant loss of the population of the already moderated, low-energy neutrons. Neutron absorption loss in $^2$H (i.e., D) is much less of an issue, so even though the cross-section is an order of magnitude smaller, D compares favorably to H as a good neutron moderator. 
Since the ground state of hydrogen forms a diatomic molecule, the scattered wavefunctions of the incident neutron off the two identical nuclei within a single molecule interfere coherently. Together with the spin-dependent nuclear force, the inter-molecular interference leads to different scattering amplitudes between states with distinct nuclear spin. 
Furthermore, spin statistics modify the molecular form factor, giving rise to selection rules governing the coupling to discrete sets of rotational states \cite{Young64}.
In fact, applying Fermi-Dirac statistics in the diatomic H$_2$ molecule was the key to solve the puzzle of discrepancies between measured values of nuclear cross-section of neutron-proton scattering and theoretical predictions using the nuclear bound state of D \cite{Schwinger37,Hamermesh46}, during the early construction of nuclear structure theory. To this end, most cross-section evaluations have properly included this coherent spin effect \cite{MacFarlane94}. 

On the other hand, all reported works (including NJOY \cite{njoy99}, \cite{liu2000}, \cite{atchison2007}, \cite{Granada09}) that evaluate inelastic cross-sections in the thermal and cold regimes for use in MCNP employ the incoherent approximation (IA) 
that assumes negligible interference between wavefunctions scattered from different lattice sites in the intra-molecular scale. In the IA, the amplitude of coherent scattering is evaluated using the same algorithm used in evaluating the incoherent scattering.  
The use of IA is popular because the inelastic incoherent cross-sections can be readily calculated via standard phonon expansions, only requiring limited information on density of states (or the frequency distribution, as referred in some literature) of the collective excitations as an input.
While IA works quite well for moderators consisting of hydrogenous materials ($\sigma_H^{coh}$=1.7583 b, $\sigma_H^{inc}$=80.27 b),
in the case of D$_2$, the contribution of the coherent scattering is not small ($\sigma_D^{coh}$=5.592 b, $\sigma_D^{inc}$=2.050 b) and thus the adoption of IA should be examined carefully. 
Especially for low energy neutrons with wavelengths comparable to and larger than the lattice constants, the use of IA seems particularly deficient. To address the shortcomings of this assumption and all previous calculations that invoked this assumption, we have developed the first full treatment of neutron scattering for neutrons with wavelength larger than angstroms. 
This calculation includes both the coherent and the incoherent scattering which excite single and multiple phonons, as well as the processes which excite rotational transitions. The elastic Bragg scattering is also included.
In this paper, we will show that this full model, when applied to solid D$_2$, significantly modifies the cross-sections of ultracold neutrons (UCN), which are free neutrons with E/$k_B<$ 3~mK, where $k_B$ is the Boltzmann constant. 

The scattering process responsible for UCN production differs from the thermalization process in the thermal and cold neutron moderators, in which multiple scatterings take place while the energy spectrum evolves continuously until the neutron gas establishes thermal equilibrium. In contrast, superthermal UCN production occurs as a single scattering event during which the incident neutron comes to a near-full stop, giving up its energy and momentum to a matching quasi-particle created in the interacting media. Even though the phase space of this inelastic scattering process is limited, the cross-section is non-negligible, provided that the interacting medium has excitations that coincide with the energy and momentum of incident cold neutrons. Subsequent scattering of the down-converted UCN (mostly upscatterings) is suppressed inside a superthermal UCN source by simply reducing the thermal population of the quasi-particles. Cooling the moderator effectively pumps out these quasi-particles. For the typical application of neutron thermalization down to a few milli-eV, scattering kernels evaluated based on IA give reasonable predictions of the neutron yields. However, many criteria of IA break down when applied to the extreme low-energy regime (E $<$ 350 neV) of UCN physics. 
Re-evaluating the UCN cross-sections in solid ortho-D$_2$ using the full model shows that UCN cross-sections differ significantly from previous estimates using the IA approach~\cite{liu2000}.

\section{Dynamics in solid D$_2$}

Calculation of the coherent inelastic scattering in solid D$_2$ requires detailed knowledge of the dynamics and the dispersion of the collective energy excitations. 
Fortunately, the dynamics of D$_2$ in a molecular lattice were studied extensively in the early days of neutron scattering experiments~\cite{Squire55, Egelstaff67, Elliott67, Schott70, Diehl75, Danchuk04, Nielsen71,Nielsen73,Schmidt84}. Due to the large lattice constants resulting from the large zero-point motion, molecular rotation remains free in the solid matrix and the translational degrees of freedom can be decoupled and the coupling strength estimated using an isotropic Lennard-Jones potential. The crystal is simple enough that the dispersion energy and the polarization vector of distinct phonon branches can be calculated with a tensor-force model, without the use of molecular dynamic simulations that require intensive numerical computations.
Dewames {\it et al.} ~\cite{Dewames65, Lehman62, Collins62} developed a Born-von Karman model, using an axially symmetric (A-S) two-body potential including up to the third nearest-neighbors, for systems with the hexagonal close packed (HCP) structure. This dynamical matrix model was applied successfully to HCP terbium \cite{Houmann70}, and recently extended to HCP Tb, Sc, Ti and Co \cite{Vaks08}. 
Using the force constants measured by Nielsen \cite{Nielsen73} in the A-S model framework, we re-constructed the full energy dispersion that is a function of the three dimensional momentum vector of phonons propagating in solid ortho-D$_2$. 
This angular-dependent dispersion function is then used in a Monte-Carlo code to construct the inelastic coherent scattering cross-section for polycrystalline solid D$_2$. 
For each scattering event sampled by the Monte-Carlo, the scattering kinematics is determined and the coherent scattering amplitude is calculated following the standard textbook formalism \cite{loveseyBook}. 
The major difference in evaluating the coherent and incoherent scattering cross-sections lies in the kinematic conditions of momentum conservation. For the incoherent process, the scattered wavefunctions are not summed coherently as the phase coherence is destroyed by fluctuations of the scattering sites. This leads to a relaxation of the momentum conservation law typically applied to two-particle systems.  
Instead the approach of IA uses the density of states of the phonon modes to estimate the relative contribution of the scattering amplitude for phonons of different energy. The energy of phonon equates to the energy transfer of the scattered neutron a result of energy conservation. As a consequence, cross-sections evaluated using IA has a diffuse smooth $Q$ dependence without any localized enhanced intensity, as would have been expected in coherent scatterings.    

Solid hydrogen and deuterium are considered quantum solids due to the light mass of individual molecules. The quantum nature is determined by its thermal wavelength, defined as $\lambda_T=2\pi M/\sqrt{3k_BT}$~\cite{loveseyBook}. For D$_2$, it is 4.6~$\mbox{\AA}$ at 5~K and 3~$\mbox{\AA}$ at 18~K.  
For a light molecule (and atom), its large thermal wavelength exdending beyond the inter-molecular spacing causes direct overlap of the wavefunctions of adjacent molecules. This could lead to quantum inteference. Provided that the adjacent molecules are identical particles, the Bose-Einstein statistics could significantly modify the collective behavior of the molecules, as in the case of superfluid helium. Here, the ortho-state of D$_2$ molecule has possible nuclear spin of 0 and 2 and distinct sub-states of $\arrowvert I,m\rangle$=(0,0), (2,2), (2,1), (2,0), (2,-1), (2,-2) that are equally populated. 
Given that the seperation between molecules on the HCP lattice is $d=3.607\mbox{\AA}$, the mean seperation of particles with identical states is $^3\sqrt{6}\times d = 6.55 \mbox{ \AA}$, which is larger than the thermal wavelength by $\sim$ 40\%. This suggests that at temperatures higher than 5~K, solid D$_2$ can be treated as a classical system to the first order approximation. Many ongoing research~\cite{Danchuk04, Colognesi2009} are looking for the evidence of quantum interference in H$_2$, HD and possibly D$_2$. 


\begin{figure}[b]
\centering
\includegraphics[width=3.5 in]{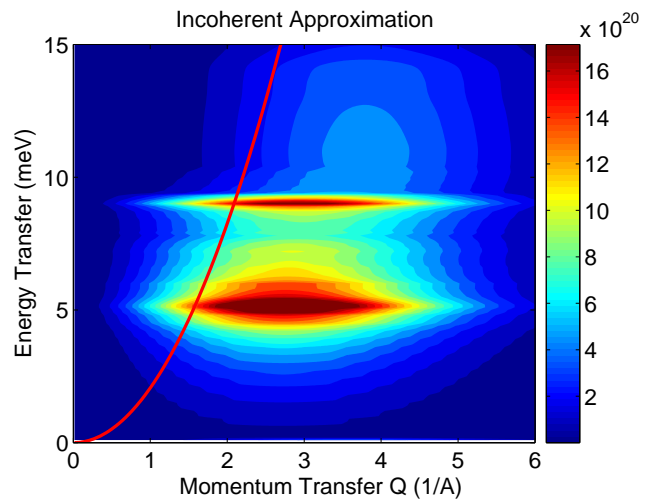}
\caption{\label{fig:SSinc} The dynamic structure function of solid ortho-D$_2$ at 5 K,  calculated using incoherent approximation (IA), presented as a contour plot. The color bar on the left indicates the intensity of the structure function. Both 1-phonon and 2-phonon processes are included. The solid curve is the free neutron energy dispersion curve, which outlines the kinematic condition for UCN production.}
\end{figure}

In a classical system, the double differential cross-section of neutron scattering is conventionally expressed as: 
\begin{equation}
\frac{d^2\sigma}{d\Omega d\varepsilon}=\frac{k}{k_0}S(\vec Q,\omega),
\label{eqn:doubleDiff}
\end{equation}
where $k/k_0$ arises from the ratio of phase space before and after scattering, and the dynamic structure function, $S(\vec Q,\omega)$, contains all the detailed physics of the neutron interactions. The dynamic structure function is related to the scattering kernel used in the neutron-moderation Boltzmann equation, in which the rate of neutron moderation is directly proportional to the dynamic structure function of the interacting medium.  
For the remaining of the paper, we examine neutron scattering in a classical solid lattice, where the scattering can be described as 
\begin{eqnarray}
S(\vec Q,\omega)&=&b^2e^{-2W} \Bigg\{\sum_{\vec G}\delta(\vec Q-\vec G)\delta(\omega)+
\sum_{j,\vec q}\frac{\hbar^2(\vec Q \centerdot \vec e_j)^2}{2M\omega_j(\vec q)} \nonumber \\
&& \Big[\sum_{\vec G} (n_j(\vec q)+1)\delta(\vec Q-\vec q-\vec G)\delta(\omega-\omega_j(\vec q)) 
\nonumber \\
&& \mbox{\hspace{0.05 in}} 
+ \sum_{\vec G} n_j(\vec q)\delta(\vec Q+\vec q-\vec G)\delta(\omega+\omega_j(\vec q)) \Big] \nonumber \\
&& \mbox{\hspace{0.05 in}} +O(Q^4) + ...\Bigg\}. \label{eqn:coh}
\end{eqnarray}
The scattering length $b$ is determined by the strength of interaction between the neutron and the scattering site. 
The Debye-Waller factor $W$ is a result of the de-localization of the atoms on the lattice sites due to thermal motion (as well as zero-point motion), smearing out the sharpness of constructive interferences of the scattered wavefunctions.
On a lattice with simple cubic symmetry, at low temperatures, $W$ reduces to
\begin{equation}
2W \rightarrow \frac{3}{4}\frac{\hbar^2Q^2}{2M\omega_D}  \qquad \mbox{as  } T\rightarrow 0,
\label{eqn:DW_cubic}
\end{equation}
where $M$ is the mass of the lattice site. For solid D$_2$, $M$ is the total mass of the D$_2$ molecule. The Debye temperature $\omega_D$ for solid D$_2$ is around 110~K.

The first term in Eq.(\ref{eqn:coh}) is the coherent elastic scattering, during which the energy transfer of the neutron is zero but the momentum transfer can be absorbed by the whole lattice if the Bragg condition ($\vec{Q}=2k\mbox{sin}(\theta/2)=\vec{G}$) is satisfied. The next term describes the inelastic scattering where the neutron energy is released to create a single phonon in the $j^{th}$ mode with energy $\omega_j(\vec q)$ and polarization vector $\vec e_j$. The occupation number of the created phonon is $n_j(q)$, which satisfies the Bose-Einstein statistics. 
The delta functions enforce energy and momentum conservation. Any change in momentum of the incident neutron during collision is picked up by a phonon of momentum $\vec q$. The total momentum conservation is adjusted by the appropriate inverse lattice vector $\vec G$ if the change of momentum is beyond the first Brillouin zone. The third term leads to energy and momentum gain by the neutron through absorption of a single phonon. 
Higher-order terms describe multi-phonon processes, with rapidly decreasing probabilities for small momentum transfer $\vec Q$.
Finally, with D$_2$ molecules arranged in a solid lattice with lattice vector $\vec a_1, \vec a_2, \vec a_3$, the inverse lattice vectors $\vec{G}$ along the principle axes are 
\begin{eqnarray}
\vec g_1 = \frac{2\pi \vec a_2 \times \vec a_3}{V}, \mbox{\hspace{0.2 in}} 
\vec g_2 = \frac{2\pi \vec a_3 \times \vec a_1}{V}, \mbox{\hspace{0.2 in}} 
\vec g_3 = \frac{2\pi \vec a_1 \times \vec a_2}{V}.
\nonumber 
\end{eqnarray}
where $V=\vec a_1 \centerdot (\vec a_2 \times \vec a_3)$ is the volume of the unit cell.

\section{Elastic Scattering and Form Factors}
\label{sec:elastic}

Even though the elastic scattering is well understood, comparing the experimental data of cross-sections with the calculation allows for consistency checks on the occasionally non-trivial molecular form factors.
This is particularly useful when calculating cross-sections in solid oxygen, where the spin form factor is less well-known than the form factors of atomic nuclei, with additional complications due to preferential alignments of the molecular axis relative to the crystal axes.
For D$_2$, the situation is somewhat simpler because the rotational motion remains free in the solid lattice, and the molecular wavefunction can be described by spherical harmonics without preferred orientations~\cite{Young64}.
The total cross-section for elastic scattering (integrating the 
first term in Eq.(\ref{eqn:coh})) can be separated into coherent Bragg part and incoherent diffuse part \cite{Egelstaff54, loveseyBook}: 
\begin{eqnarray}
\sigma^{coh}_0&=&4\pi\frac{\lambda^2}{8\pi V}\sum_{hkl}d_{hkl}|F^{coh}_{hkl}|^2e^{-2W} \mbox{, and} \label{eqn:cohelastic} \\
\sigma^{inc}_0&=&4\pi |F^{inc}|^2 e^{-2W},
\end{eqnarray}
where $d_{hkl}$ is the inter-planar spacing between the [hkl] planes in the lattice space, i.e.,
\begin{equation}
d_{hkl}=\frac{2\pi}{|\vec G|}, \mbox{\hspace{0.2 in}where   } \vec G= h \vec g_1 + k \vec g_2 + l \vec g_3.
\end{equation}
The form factor $F$ within the unit cell differs depending on whether the coherence of the scattered wave is preserved after scattering. For coherent processes,  
\begin{eqnarray}
F^{coh}_{hkl} &=&\sum^{N_{basis}}_{i=1}b^{coh}_i e^{i \vec G \centerdot \vec R_i} \mbox{\hspace{0.2 in}  if }\lambda \geqslant 2d_{hkl} \nonumber \\
&=& 0 \mbox{\hspace{1 in}  otherwise.}  
\end{eqnarray}
For incoherent processes, 
\begin{eqnarray}
F^{inc} &=&\sqrt{\sum^{N_{basis}}_{i=1} (b^{inc}_i)^2} \mbox{ .} \hspace{1 in} 
\label{eqn:Finc}
\end{eqnarray}
In a non-Bravais crystal, the number of atoms in the unit cell, $N_{basis}$, is larger than 1. 

On each lattice site in the D$_2$ crystal resides a diatomic molecule, the additional molecular form factor of which significantly modifies the scattering length. For D$_2$, the bosonic nature of the two identical deuteron nuclei leads to unique couplings to a set of molecular rotational states of either even or odd symmetry, so that the symmetry of the whole system is preserved under permutation of identical particles.   
It follows that there exist two different types of D$_2$ molecules, i.e., ortho-D$_2$ with even nuclear spin ($I$=0, 2) coupling to symmetric molecular rotational wavefunctions (with orbital angular momentum $J$=0,2,4,...), and para-D$_2$ with odd nuclear spin ($I$=1) coupling to anti-symmetric molecular wavefunctions (of $J$=1,3,5,..). 
By convention the ortho-state designates the group with the highest spin multiplicity.
On account of the spin-dependence of the nuclear force, neutron scattering lengths of ortho and para-D$_2$ are different. In addition, when interacting with one species of diatomic molecule, say ortho-D$_2$ molecules (with $I$=0,2 and $J$=0 in the ground state), the neutron scattering length also depends on the final state of the target molecule \cite{Young64}, i.e., 
\begin{eqnarray}
b_{D^2}^2\Big\vert_{J=0\rightarrow0}&=&\Big(a^2_{coh}+\frac{5}{8}a^2_{inc}\Big) C^2(000;00)|A_{00}|^2 \mbox{,\hspace{0.2 in}} \label{eqn:J00} \\ 
b_{D^2}^2\Big\vert_{J=0\rightarrow1}&=&\frac{3}{8}a^2_{inc}(2\cdot 1+1) C^2(011;00)|A_{01}|^2 \mbox{,  } \\
b_{D^2}^2\Big\vert_{J=0\rightarrow2}&=&\frac{3}{8}a^2_{inc}(2\cdot 2+1) C^2(022;00)|A_{02}|^2 \mbox{,  } \label{eqn:J01} 
\end{eqnarray}
where $a_{coh}=6.67$~fm and $a_{inc}=4.04$~fm, and the molecular form factors are: 
\begin{equation}
A_{00}=2j_0\Big(\frac{Q a}{2}\Big), \hspace{0.1 in} A_{01}=2j_1\Big(\frac{Q a}{2}\Big), \hspace{0.1 in} A_{02}=2j_2\Big(\frac{Q a}{2}\Big),
\end{equation}
with $j_l$ as the spherical Bessel function of order $l$. The separation of the two deuteron nuclei is $a=0.71\mbox{ \AA}$. All Clebsch-Gordan coefficients used in the above equations are 1. The $J=0\rightarrow1$ process resulting in the ortho- to para- transition requires that the neutron to transfer energy equal to the rotational transition energy, and thus it is only present in the inelastic scattering. 

\begin{figure}
\centering
\includegraphics[width=3.2 in]{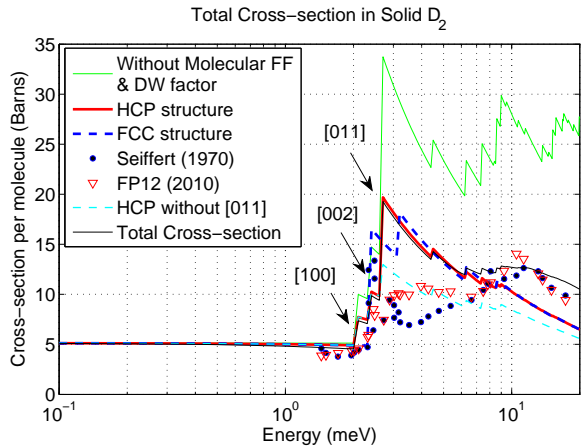}
\caption{\label{fig:elastic} Elastic total cross-sections and experimental data reported by Seiffert \protect\cite{seiffert1970}(solid circles) and our measurements\protect\cite{Lavelle2010} (triangles). Inclusion of the molecular form factor and the Debye-Waller factor shifts the polycrystalline cross-section prediction from the solid green curve to the solid red curve. Elastic cross-sections for both polycrystalline HCP (solid red) and FCC (dashed blue) lattice are plotted for comparison.}
\end{figure}

The energy dependence of the total elastic scattering cross-section of polycrystalline samples is determined by the static structure function of single-crystal, which is HCP for D$_2$ \cite{Bostanjo67}. However, some Raman scattering data \cite{Collins96,Stein72} indicate an FCC component depending on the temperature and the pressure of the solid.
Fig.~\ref{fig:elastic} shows the calculations together with experimental data.
Note the molecular form factor and the Debye-Waller factor together significantly reduce the elastic scattering amplitude, as both suppress the neutron scattering amplitude associated with large momentum transfers (see Fig.~\ref{fig:DW} and more discussion in Sec.\ref{sec:inel}).
Note also that the experimental data do not show the large diffraction peak due to scattering off the [011] plane. There is evidence that the solid D$_2$ grown in a cold cell tends to self-anneal, becoming a few single crystals of large size. In these experiments, neutrons scatter through diffractions from several single crystals of large size oriented at different angles. 
As a result, the total cross-section deviates from the prediction in the powder limit described by Eq.(\ref{eqn:cohelastic}). 

The presented total cross-section is extracted from neutron transmission measurements. Excluding the [011] scattering alone brings the prediction on top of the experimental data (see the light blue dashed curve in Fig.~\ref{fig:elastic}).
In these measurements, the finite area of the neutron detector would lead to reduced scattering amplitudes, because neutrons scattered at low angles are still detected.
However, the surface coverage of the detector has to be unphysically large in order to account for the missing amplitude. This rules out detector geometry as the cause of the missing scattering amplitude.
On the other hand, missing the major [011] peak suggests that the crystal is grown preferably with the c-axis perpendicular to the direction of the neutron beam. This would lead to reduced Bragg scattering for small momentum transfer centered around the direction of the c-axis.
The reduced scattering cross-section could also be explained by multiple scattering processes, in which some neutrons that are already scattered once scatter again back into the beam and thus enhances the neutron transmission. The presence of multiple scattering leads to a reduction in the cross-section if the analysis algorithm does not include a correction for multiple scattering.
Due to this complication, the data of neutron transmission, which was collected along with the data of UCN production, cannot be used to distinguish between the HCP and FCC lattice structures.
Finally, the discrepancy between the experimental data and the theoretical predictions at neutron energies higher than 10~meV is accounted by including contributions from inelastic scattering, as explained in the following section.

\section{Inelastic Scattering}
\label{sec:inel}

\begin{figure}[!t]
\centering
\includegraphics[width=3.5 in]{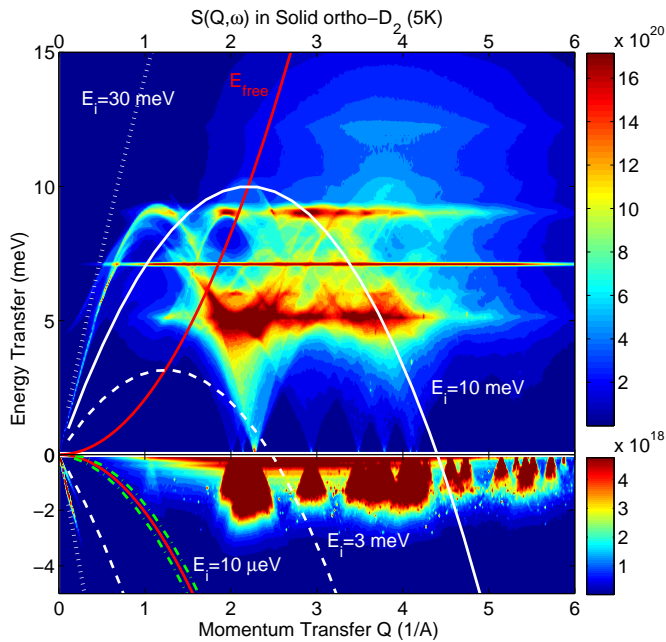}
\caption{\label{fig:SSDT}A contour plot of the dynamic structure function, S(Q,$\omega$), of solid ortho-D$_2$ at 5~K, plotted for both neutron downscattering (positive $\omega$) and neutron upscattering (negative $\omega$). To better resolve the details of the phonon modes, the upscattering region is plotted on an amplified scale compared to the downscattering region. The red curves are free neutron energy dispersion curves, outlining the allowed kinematics for UCN production and UCN upscattering. The light curves form the kinematic bounds for incident neutrons with initial energy 30~meV (fine white dots), 10~meV (white solid), 3~meV (white dashed), and 10~$\mu$eV (green dashed-dot).}
\end{figure}

The inelastic scattering involving creations and annihilations of single phonon in Eq.(\ref{eqn:coh}) is calculated numerically for coherent and incoherent processes with corresponding scattering lengths given by Eqs.(\ref{eqn:J00}) and (\ref{eqn:J01}). 
To evaluate coherent scattering in a polycrystalline sample beyond the IA prescription, we developed a Monte-Carlo algorithm to sample the dispersion curve of each single crystallite oriented at random angles, strictly following the kinematics of energy and momentum conservation in each coherent scattering event \cite{Liu2004}. 
Taking the angular average of the momentum transfer vector simplifies the orientation-dependent scattering amplitude into a two dimensional map of S(Q, $\omega$).
The result of one such calculation (a high resolution map with 500$\times$200 grids in the (Q, $\omega$) space) is shown in Fig.~\ref{fig:SSDT}.
The calculation includes 1-phonon and 2-phonon contributions from both the coherent and incoherent process, calculated independently for the ortho $\rightarrow$ ortho ($J=0\rightarrow0$) and the ortho $\rightarrow$ para ($J=0\rightarrow1$) transition. 
Note that the neutron scattering resulting in ortho $\rightarrow$ para transition involves spin flip, and thus is a purely incoherent process. The relative weight of the scattering amplitude of the ortho-ortho and ortho-para processes is dictated by Eqs.(\ref{eqn:J00}) and (\ref{eqn:J01}). In addition, the calculation also includes inelastic scattering leading to the rotational excitation $J_{01}$, where incident neutrons lose energy in ortho-D$_2$ through inter-molecular excitations without activating translational phonons via a nuclear recoil. In upscattering, this $J=0\rightarrow 1$ process is absent in ortho-D$_2$ for neutrons with energy less than the $J_{01}$ transition energy of 7.1~meV.   

The average algorithm used in polycrystalline samples prescribes that, for every given magnitude of momentum transfer, a random angle is assigned to determine the momentum transfer vector. Setting this momentum transfer to that of the phonon, the eigenfrequencies and eigenvectors are calculated uniquely using the force tensor matrix. For the HCP structure, the result consists of six eigenvalues, each corresponding to a translational degree of freedom originated from the two molecules in a unit cell. The energy transfer bin that contains the calculated eigen-frequency is then augmented by a numerical value of the scattering amplitude estimated using Eq.(\ref{eqn:coh}). For each magnitude of momentum transfer, up to 10$^5$ different angles are sampled isotropically over $4\pi$. 
In the end, a two-dimensional map of the dynamic structure function is evaluated for positive energy transfer, which corresponds to neutron downscattering. A second map is evaluated independently for negative energy transfer (neutron upscattering). The main difference between upscattering and downscattering arises from the dependence on the phonon occupation number $n$. Upscattering occurs when the neutron encounters a phonon, and thus the scattering probability scales with the thermal population of phonons. In contrast, downscattering of neutrons through phonon creation takes place only when the kinematics of energy and momentum transfer is favorable, as determined by the existing modes of phonon excitation. The downscattering amplitude scales as (1 + $n$), where $n$ weakly enhances the transition probability due to the bosonic nature of phonons.   

\begin{figure}[!t]
\centering
\includegraphics[width=3.0 in]{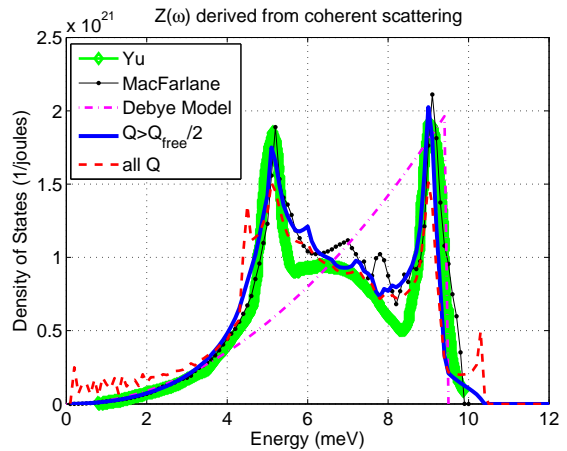}
\caption{\label{fig:ZZ} The density of states, $Z(\omega)$, derived from the coherent dynamic structure function S$^{coh}$(Q, $\omega$) in solid ortho-D$_2$ at 5~K, with different lower bounds of the integration range in $Q$. The results are compared to the density of states presented in Yu \cite{Yu86} (solid green line), MacFarlane~\cite{MacFarlane94} (solid black line with dots), and that of a simple Debye model (magenta dashed-dot line).}
\end{figure}

Next, to evaluate the incoherent contributions, we construct the density of states by integrating the coherent dynamic structure function derived in the previous step, after correcting all the dynamical factors, over the whole accessible range of momentum transfer: 
\begin{equation}
Z(\omega)=\int_{Q=\frac{1}{2}Q_{free}}^{\infty} \frac{\omega}{\hbar^2Q^2/2m}\frac{e^{2W}}{(n(\omega)+1)}S^{coh}(Q,\omega) dQ.
\end{equation} 
However, simply integrating over all possible $Q$ values leads to a non-vanishing density of states as $Q$ approaches zero. This disagrees with the Debye model, which prescribes the density of states for phonons as $cQ^2$, where $c$ is the speed of sound in the solid.
To fix this discrepancy, we set the lower bound of the integration to half of the momentum carried by a free neutron, for the following reasons:
As illustrated in Fig.~\ref{fig:SSDT}, the scattering amplitude peaks beyond the first Brillouin zone. The accessible kinematic range of free neutrons in the cold regime (with incident energy less than 10~meV) simply cannot produce excitations within the first Brillouin zone, except for those close to the zone boundaries. In other words, the sound speed of phonons exceeds the group velocity of the neutron wave-packet, leading to no coupling between neutrons and phonons of low momentums. Excluding the rather large amplitude for coherent creation of acoustic phonons with momentum smaller than half of the free neutron momentum (in the first Brillouin zone) restores the expected $Q$ dependence at low momentum transfers (as shown in Fig.\ref{fig:ZZ}).  
Only the coherent $J_{00}$ process is included in evaluating $Z(\omega)$.

We have also calculated $Z(\omega)$ using the classical approach by integrating the derivative of the dispersion curve over the hypersurface with constant $\omega$, following a method described in~\cite{Rabubenheimer1967}. The result is consistent with the method discussed above. 
With $Z(\omega)$, the incoherent scattering law can be calculated using the standard textbook formula, without the need for the rather computation-intensive algorithm developed for the coherent process. 

\begin{figure}
\centering
\includegraphics[width=3.0 in]{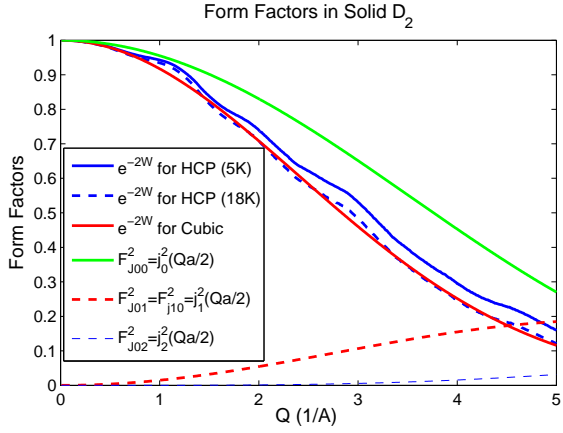}
\caption{\label{fig:DW} The Debye-Waller factor, e$^{-2W}$, and the molecular form factors for a few lowest rotational transisions.}
\end{figure}

For a final consistency check, using the derived density of states, we can calculate the Debye-Waller factor using the polarization vectors in the HCP structure~\cite{loveseyBook}:
\begin{equation}
W(Q)=3\int_0^{\omega_m} d \omega Z(\omega)\frac{\hbar^2}{4M\omega}(2n(\omega)+1)|\vec{Q}\cdot\vec{e}|^2_{av},
\end{equation}
\noindent and examine how well the approximation of the cubic symmetry of Eq.(\ref{eqn:DW_cubic}) applies to the HCP structure. We found that the Debye-Waller factor of solid D$_2$ at 18~K agrees remarkablly well with the simple cubic approximation, and that at 5~K is slightly above the value of the cubic structure (see Fig.\ref{fig:DW}). For the $Q$ range relevent to UCN production (Q $<2.5 \mbox{/\AA}$), the agreement is within 5\%. 
For completeness, the molecular form factors are plotted to compare with the Debye-Waller factor. For the $J=0\rightarrow 0$ transition, the form factor suppresses the high momentum transfer, similiar to the effect of the Debye-Waller factor. For the $J=0\rightarrow 1$ and $J=0\rightarrow 2$ transitions, the form factors start at zero and grow with increasing $Q$. Processes involving the $J=0\rightarrow 2$ transition (E$\geq$15~meV) are highly suppressed in magnitudes and thus we don't expect it to contribute significaly to the cross-section. 

After combining independent results for coherent and incoherent cross-section, the major features of the inelastic scattering are clearly visible in the dynamic structure function evaluated with our full model (see Fig.~\ref{fig:SSDT}). The acoustic phonon branches that extend from elastic Bragg peaks into high energy transfers can be clearly identified.
Only phonons created along the symmetry axes have energy dispersion that is periodic in momentum; phonons created along other directions have less symmetry, but their energy can be calculated analytically using the force tensor.
We find that the clustering of the scattering amplitude around 5~meV is due to mostly the phonon modes in the basal plane. Including scatterings with directions out of the basal plane, the phase space increases resulting in large scattering amplitude associated with these in-plane phonon modes.
The narrow band of large scattering amplitude around 9~meV is due to the longitudinal optical branch of phonons associated with scatterings along the c-axis. Multi-phonon contribution increases the scattering amplitude for energies larger than 10~meV and momentum transfer larger than 3/$\buildrel _\circ \over {\mathrm{A}}$. 

Overall, the scattering amplitude is significantly suppressed for momentum transfers larger than 5/$\buildrel _\circ \over {\mathrm{A}}$, owing to the combined effect of the molecular form factor and the Debye-Waller factor. The same suppression due to these form factors was observed in the elastic scattering as discussed in the previous section.
In contrast to the interpretations given in \cite{Gutsmiedl2009}, our study shows that the large scattering amplitude cannot be simply associated with phonons on the zone-boundaries along the symmetry axes, because the phase space for interaction with these phonons is very small compared with that available for inelastic scattering. Finally, the pure $J_{01}$ transition (without phonon creation) is visible at a fixed energy transfer of 7.1~meV. The molecular form factor $2j_1(Qa/2)$ dictates the $Q$ dependence of the scattering amplitude in the $J_{01}$ transition, as it increases from zero with increasing $Q$. By comparison, the dynamic structure function calculated in the IA approach (as shown in Fig.~\ref{fig:SSinc}) is grossly over-simplified.



\begin{figure}[t!]
\includegraphics[height=1.6 in,width=3 in]{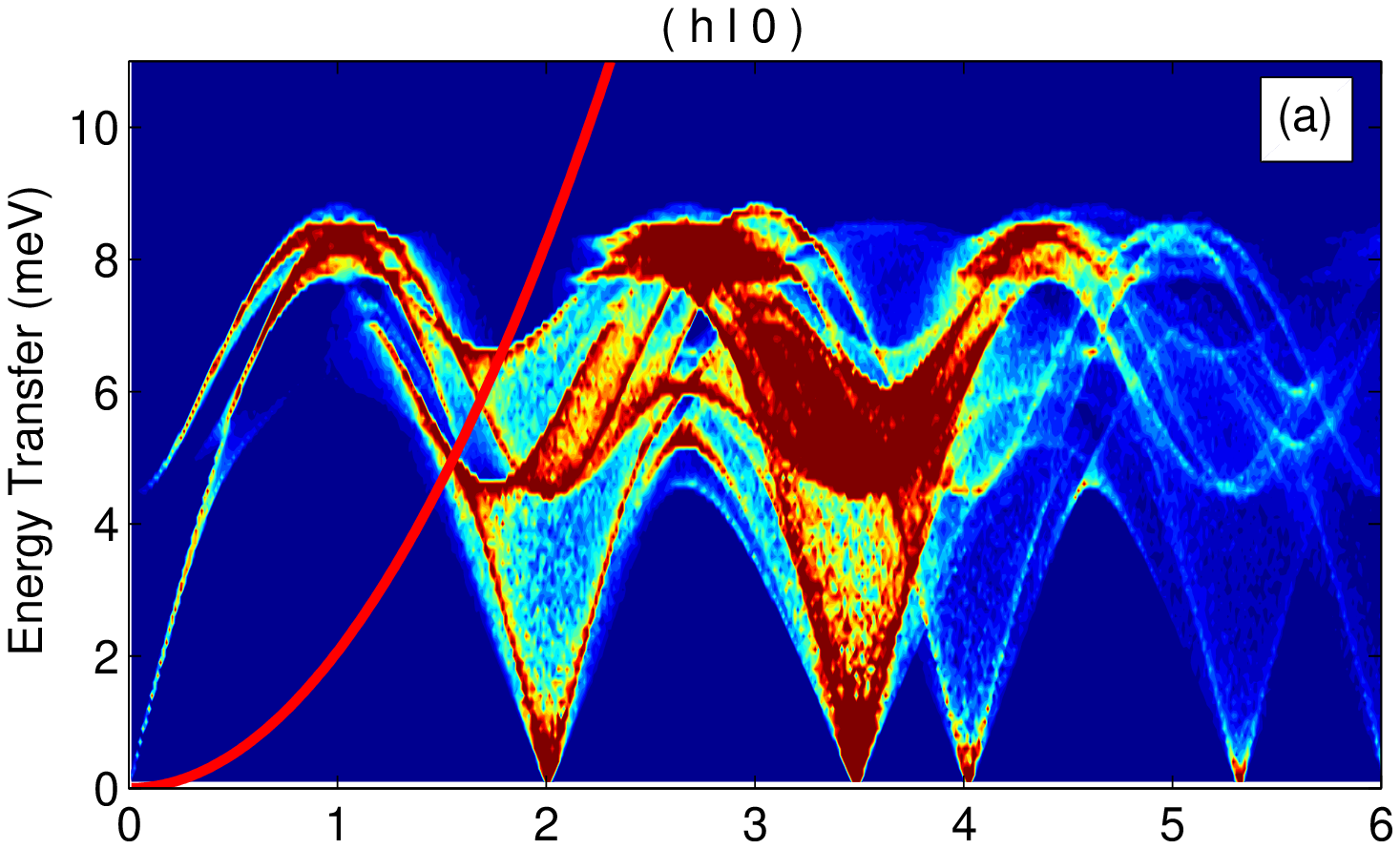}
\includegraphics[height=1.6 in,width=3 in]{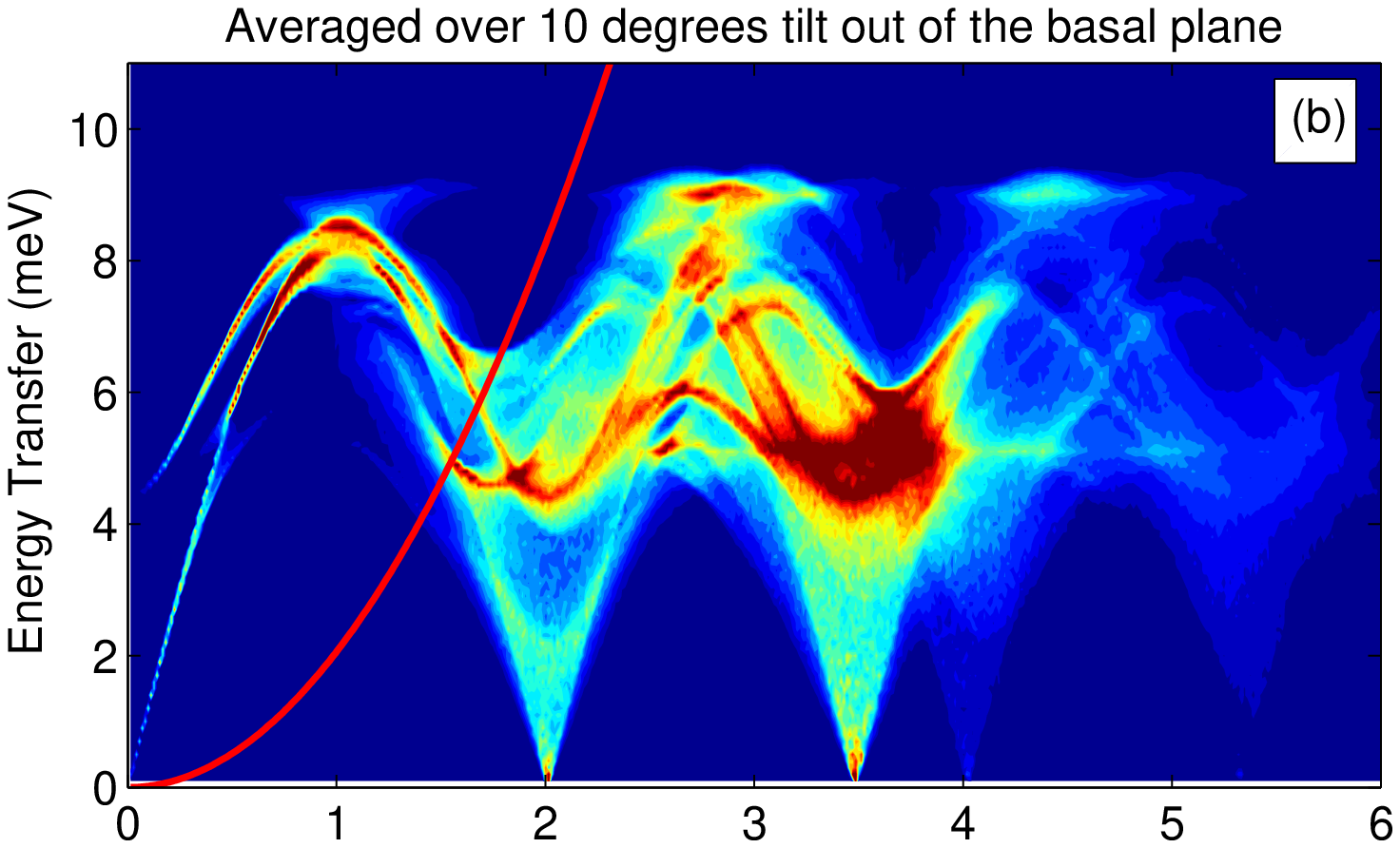}
\includegraphics[height=1.6 in,width=3 in]{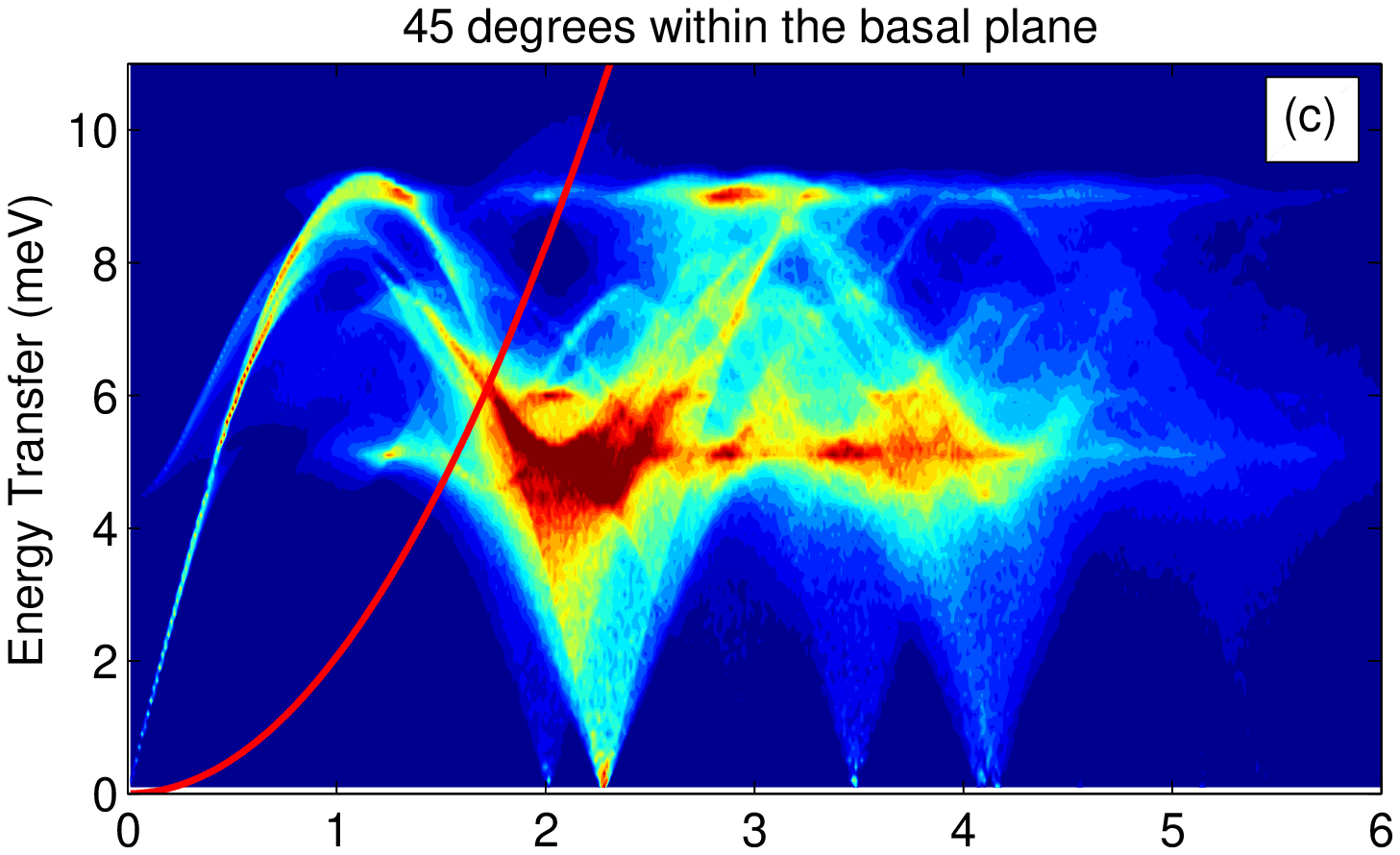}
\includegraphics[height=1.6 in,width=3 in]{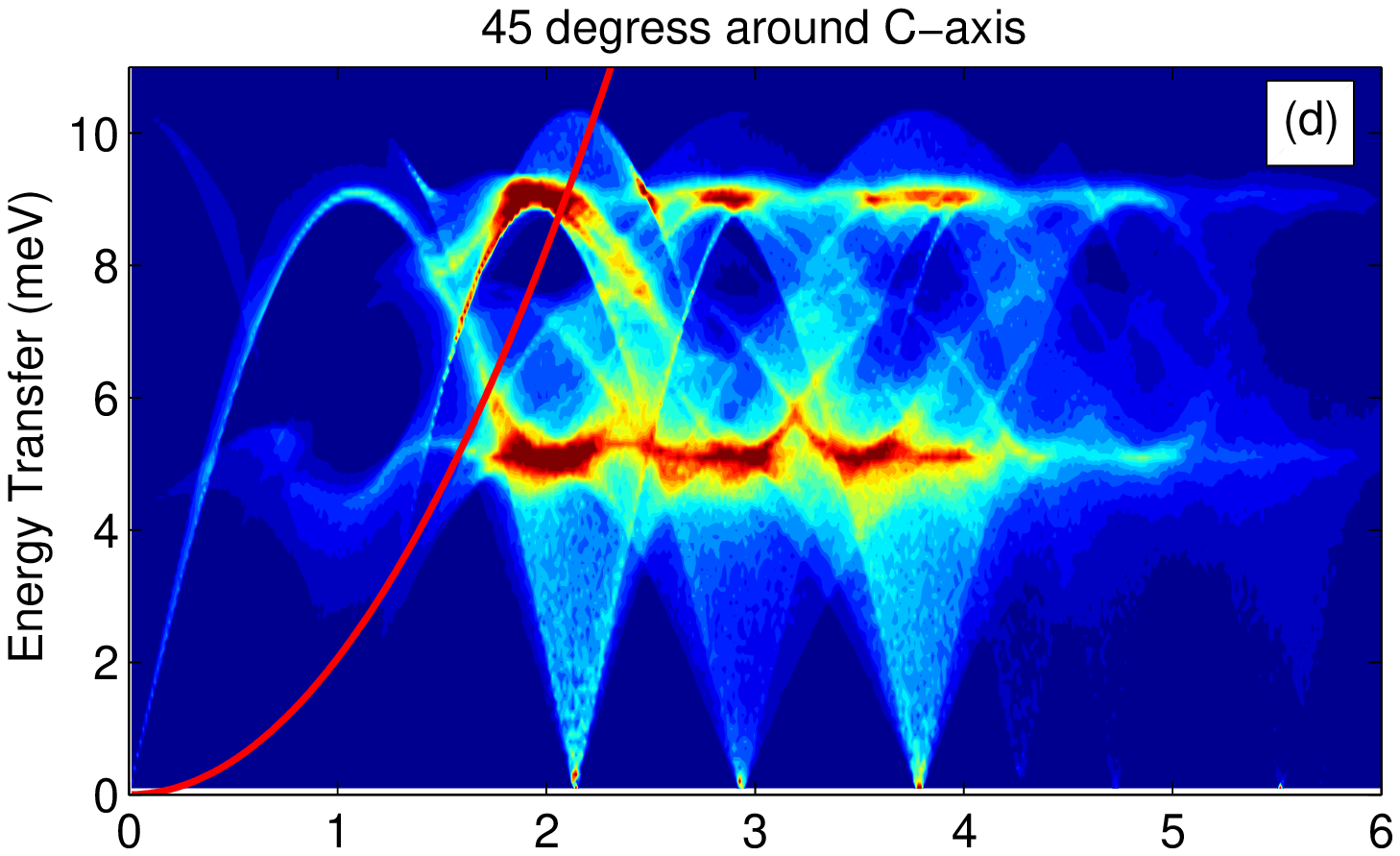}
\includegraphics[height=1.6 in,width=3 in]{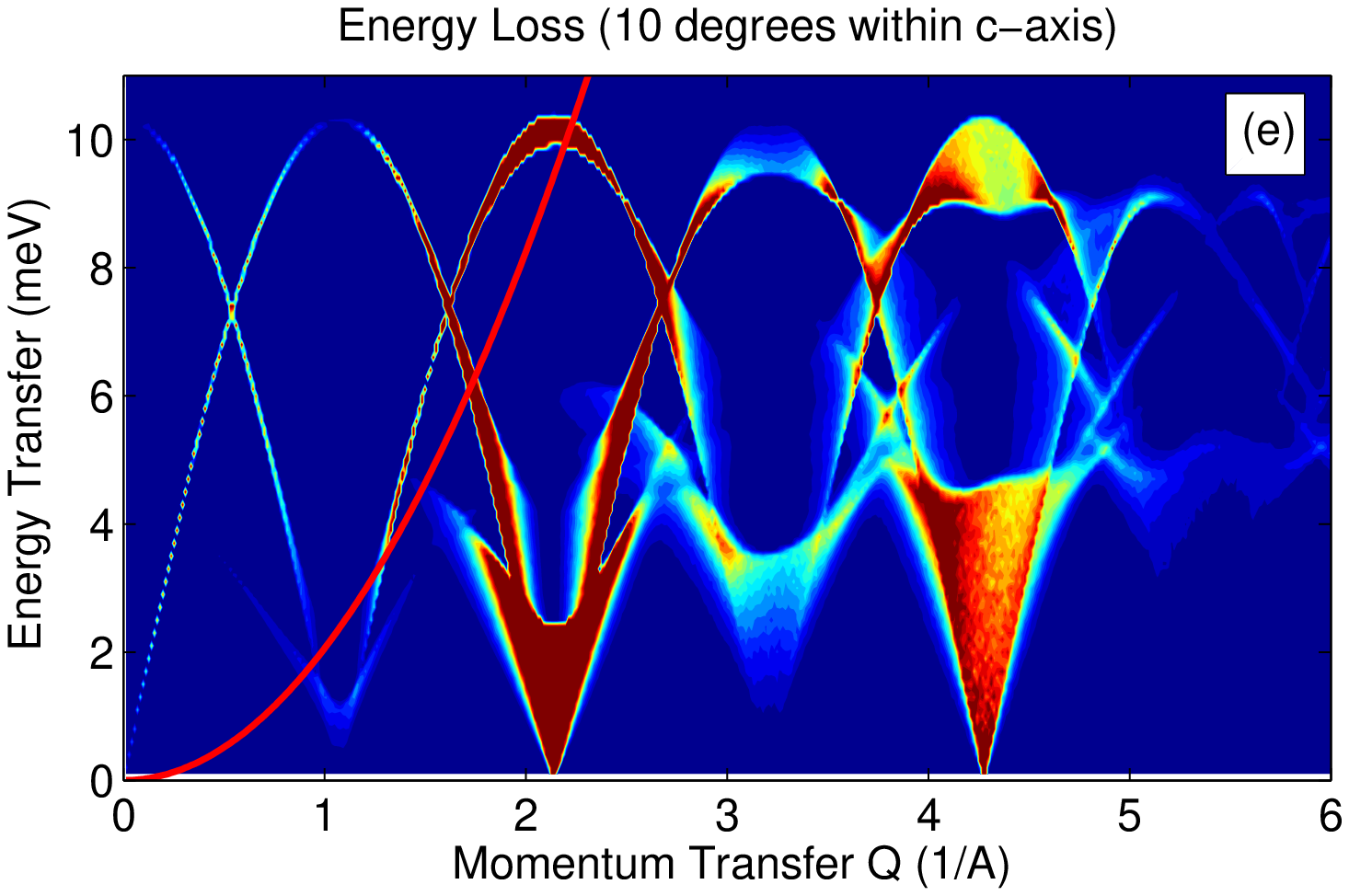}
\caption{\label{fig:crystal}The dynamic structure function S$^{coh}$(Q, $\omega$) in solid ortho-D$_2$ at 5~K. The crystal orientations in these polycrystalline samples are limited to 10$^{\circ}$ and 45$^{\circ}$ relative to the basal plane or to the long axis (c-axis).}
\end{figure}

In order to account for the scattering amplitude beyond the energy of the most energetic phonon available (i.e., the transverse optical phonon moving in the [001] direction), we have also calculated the 2-phonon contributions. 
The coherent dynamic structure function involving two phonons can be expressed as:
\begin{align}
&\frac{1}{2}\sum_{j,\vec q, j', \vec q'}\frac{\hbar^2 (\vec Q \cdot \vec e_j )^2}{2M\omega_j(\vec q)}
\frac{\hbar^2 (\vec Q \cdot \vec e_{j'})^2}{2M\omega_{j'}(\vec q')}
\sum_{\vec G} \Big[ (n_j(\vec q)+1) \nonumber \\
&\times (n_{j'}(\vec q')+1) \delta(\vec Q -\vec q-\vec q'-\vec G)\delta(\omega-\omega(\vec q)-\omega(\vec q')) \nonumber \\
&+(n_j(\vec q)+1)n_{j'}(\vec q')\delta(\vec Q -\vec q+\vec q'-\vec G)\delta(\omega-\omega(\vec q)+\omega(\vec q')) \nonumber \\
&+n_j(\vec q)(n_{j'}(\vec q')+1)\delta(\vec Q +\vec q-\vec q'-\vec G)\delta(\omega+\omega(\vec q)-\omega(\vec q')) \nonumber \\ 
&+n_j(\vec q)n_{j'}(\vec q')\delta(\vec Q +\vec q+\vec q'-\vec G)\delta(\omega+\omega(\vec q)+\omega(\vec q')) \Big],
\label{eqn:2phonons}
\end{align}
\noindent where the first term describes the creation of two phonons through a single scattering process, the second and the third term describe the simultaneous creation of one phonon and annihilation of another phonon of different energy, and the final term describes the annihilation of two phonons in a single scattering event. The combined amplitude is smaller than that of the single phonon process. Nevertheless, the two-phonon process enlarges the scattering phase space to include those at higher energy transfers, and thus neutrons with energies higher than the Debye temperature participate in neutron downscattering through multi-phonon processes.

In constructing the $S(Q,\omega)$ for the two phonon process, we implement the following steps:
For a given magnitude of momentum transfer $Q$, first throw the dice (i.e., the random number generator in the Monte-Carlo code) to determine one angle. This random angle together with $Q$ determines the vector of momentum transfer. Throw the dice again to determine the magnitude of the momentum of the first phonon, and again to determine the angle of this momentum vector. Once the two vectors are specified, the momentum of the second phonon is determined by the momentum conservation law. Now with the momentum vectors specified, we can calculate the energy of each of the two phonons and sum them up as the total energy transfer. The energy bin corresponding to the total energy transfer for this scattering event is incremented by the amplitude described in Eq.(\ref{eqn:2phonons}).  
Note that the momentum conservation and energy conservation are strictly applied in every scattering event as dictated by the delta functions in Eq.(\ref{eqn:2phonons}).  

To address the effect of preferred directions of crystal orientation as indicated by the total cross-section data (see Sec.\ref{sec:elastic}), we have also calculated the scattering laws by restricting the scattering angle relative to the inverse lattice vectors. A few examples of the resulting dynamic structure functions are shown in Fig.~\ref{fig:crystal}. To accentuate the differences, only the coherent scattering contributions are plotted.  
Comparison with recent measurements \cite{Gutsmiedl2009} suggests that the experimental data are best described by scatterings confined to within 45$^{\circ}$
of the basal plane. The detailed information contained in these inelastic scattering maps strongly suggests that the missing peak in the total cross-section (Fig.\ref{fig:elastic}) could be explained by a crystal orientation where the c-axis is aligned preferentially perpendicular to direction of neutron propagation.

\begin{figure}[t!]
\centering
\includegraphics[width=3.0 in]{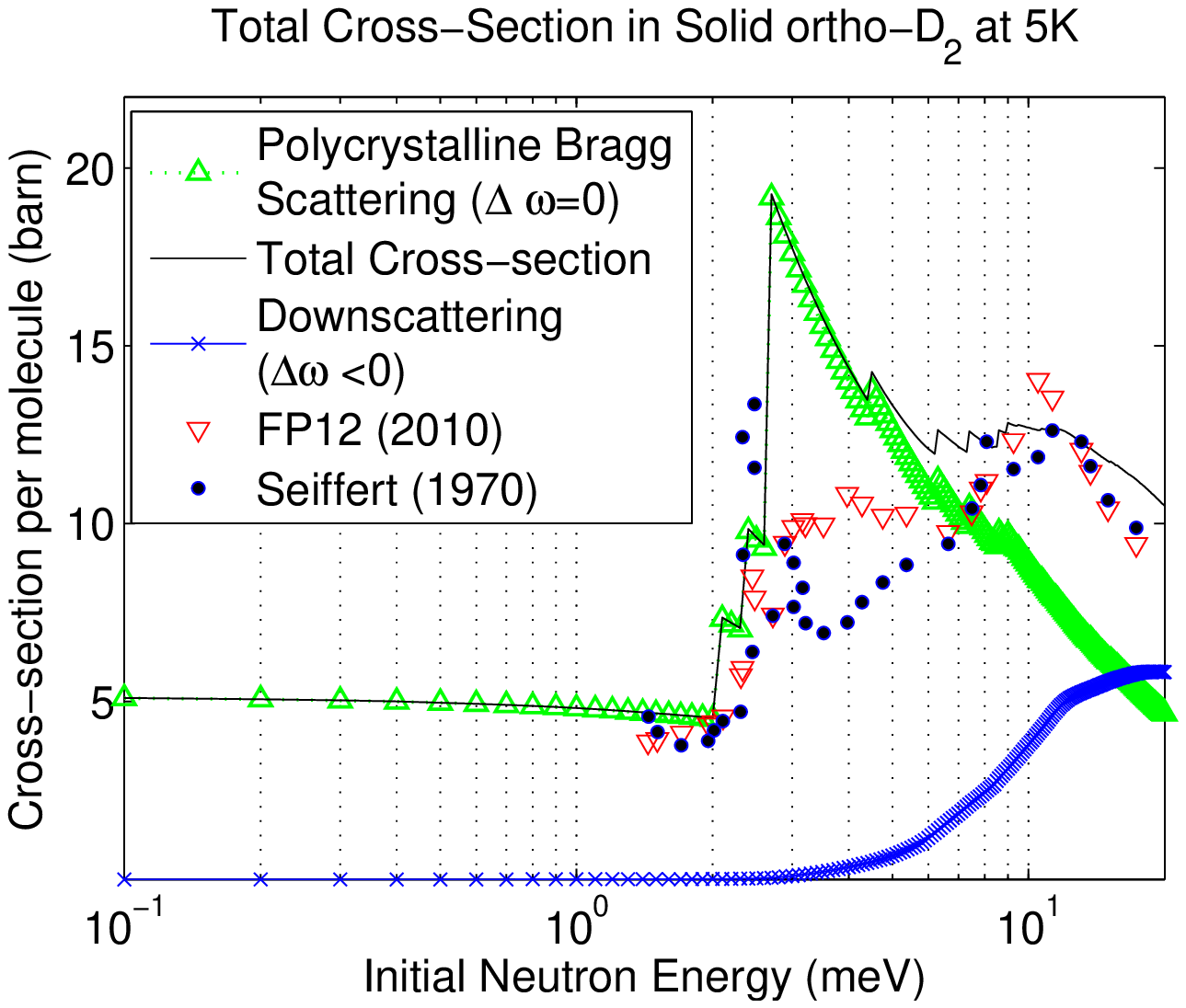}
\includegraphics[width=3.0 in]{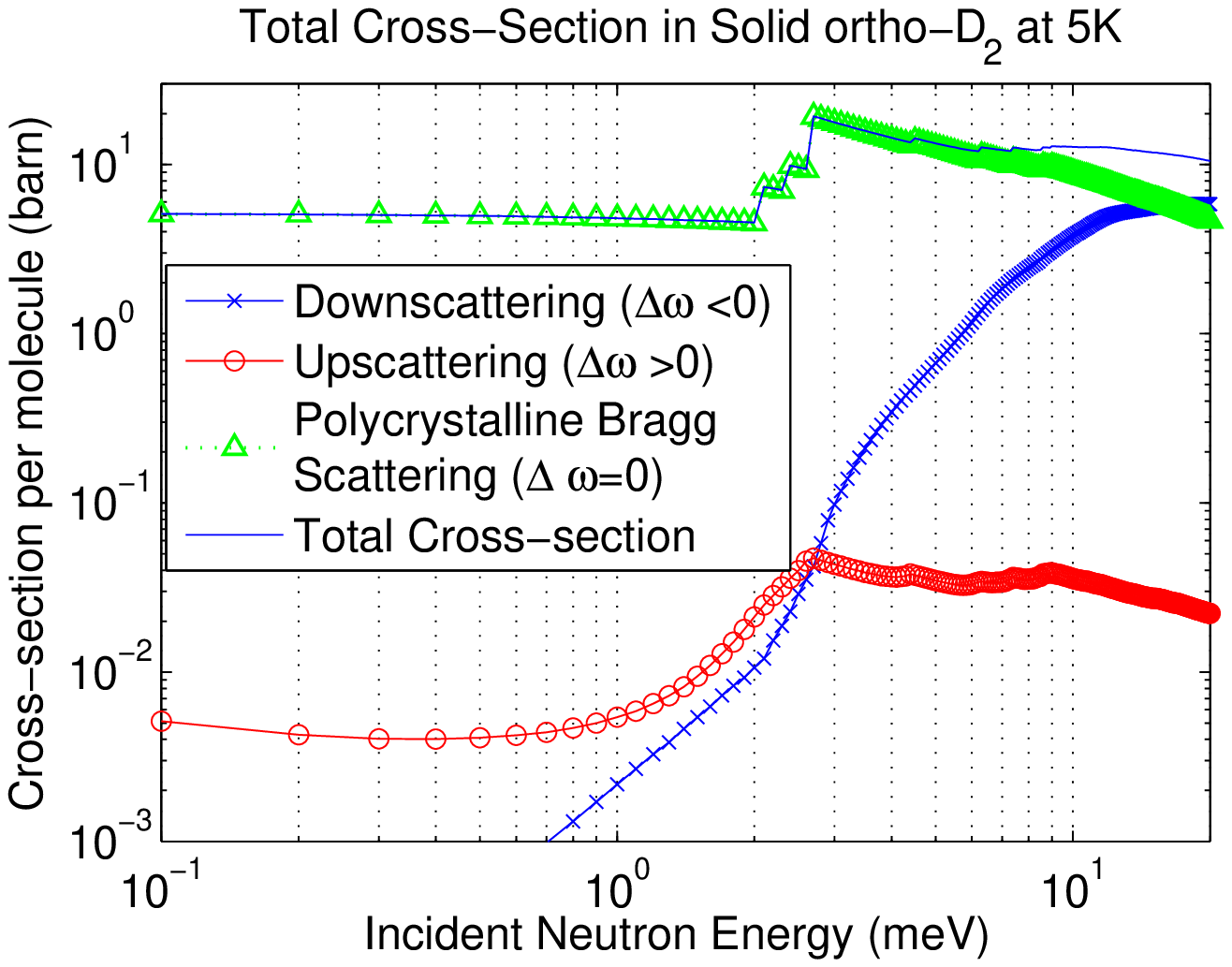}
\caption{\label{fig:SigmaTot}The calculated total cross-section per D$_2$ molecule in solid ortho-D$_2$ at 5~K, plotted in linear scale (upper graph) and log scale (lower graph). The results are compared to experimental data. The total cross-section is a sum of the elastic cross-section (green triangles), the downscattering (blue $\times$) and the upscattering (red open circles) cross-sections.}
\end{figure}

With the detailed dynamic structure function, calculation of the total cross-section becomes a straightforward integration of Eq.(\ref{eqn:doubleDiff}) over the appropriate phase space:
\begin{eqnarray}
\sigma^{tot}(E_0)&=&\frac{1}{2}\int d\Omega \int d\omega \frac{k}{k_0} S(Q,\omega) \nonumber \\
&=& \frac{1}{2k_0^2}\int d\phi \int d\omega \int_{Q_{lower}}^{Q_{upper}} dQ Q S(Q,\omega). \hspace{0.2in}
\label{eqn:int}
\end{eqnarray} 
For neutron downscattering, $S^{down}(Q, +\omega)$ (upper graph in Fig.\ref{fig:SSDT}) is integrated over the energy transfer $\omega$ from 0 to $E_0$, where $E_0$ is the energy of incident neutrons. For neutron upscattering, $S^{up}(Q, -\omega)$ (lower graph in Fig.\ref{fig:SSDT}) is integrated over energy transfer from 0 to $\infty$. For every $\omega$, the range of integration on $Q$ is bounded by the momentum transfer in forward scattering and backward scattering:
\begin{eqnarray}
|Q_{lower}|&=&\sqrt{\frac{2m}{\hbar^2}} \Big|\sqrt{E_0-\omega}-\sqrt{E_0}\Big| \nonumber \\
|Q_{upper}|&=&\sqrt{\frac{2m}{\hbar^2}} (\sqrt{E_0-\omega}+\sqrt{E_0}).
\end{eqnarray} 
The range of integration is illustrated by the white lines in Fig.\ref{fig:SSDT} for several incident neutron energies.
Since the dynamic structure function derived for polycrystalline target is already angle-averaged, the integration in Eq.(\ref{eqn:int}) over the azimuthal angle simply yields $2\pi$. In the end, to report the total inelastic scattering cross-section per molecule, the result is divided by the number of molecules contained in each unit cell, which is two in HCP D$_2$ (shown as the pre-factor of 1/2 in Eq.(\ref{eqn:int})). The same normalization is applied to the incoherent component. According to Eq.(\ref{eqn:Finc}), the lattice form factor is $(F^{inc})^2=2$. The form factor together with the normalization brings the incoherent amplitude back to the scattering cross-section of one particle. The total cross-section is plotted in Fig.~\ref{fig:SigmaTot}. For solid ortho-D$_2$ at 5 K, the upscattering is negligible for cold neutrons over the plotted energy range from 0.1 to 20~meV, whereas the downscattering amplitude becomes appreciable for incident neutron energy larger than 5~meV. Excluding the missing [011] Bragg peak ranging from 3 to 6~meV, the inclusion of contributions from inelastic scattering produces reasonable agreements to the total cross-section data derived from simple transmission measurements.   

\section{UCN cross-sections}

The cross-sections for UCN production and UCN upscattering can be evaluated simply by finding the value of $S(Q, \omega=E_{free})$ where $E_{free}=\hbar^2Q^2/2m$, which is the dispersion curve of free neutron (the parabolic curve shown in Fig.~\ref{fig:SSDT}). 
Here, the integration of Eq.(\ref{eqn:int}) simplifies because the energy scale of UCN ($<$ 350~neV) is several orders of magnitude smaller than the energy scale of cold neutrons ($\sim$1~meV). 
For inelastic scattering events leading to UCN productions, the momentum transfer $Q=k_i-k_{ucn}$ equals the initial momentum of the neutron $k_i$ due to the negligible $k_{ucn}$. Similarly, the energy transfer $\omega=E_{i}-E_{ucn}\approx E_{i}$.
In estimating the UCN production cross-section, the integration of the double differential cross-section Eq.(\ref{eqn:doubleDiff}) becomes: 
\begin{equation}
\sigma(E_i\rightarrow E_{ucn})=\frac{1}{2}\int dE_{ucn} \int d\Omega_f \frac{k_{ucn}}{k_i} S^{down}(Q,+\omega). \nonumber 
\end{equation}
A change of integration variable $dE_{ucn}d\Omega_f=dE_{ucn}d\phi d\mu = d\phi \frac{Q}{k_ik_{ucn}}d\omega dQ $, where $\mu = \cos (\theta)$, further simplifies the above integral to
\begin{equation}
2 \pi \frac{\hbar^2}{2m_nE_i}\int d\omega \int dQ Q S^{down}(Q,+\omega). \nonumber 
\end{equation}
For scattering with the final state of interest, i.e., UCN with energy $E_{ucn}$, the range of integration is a narrow band around the parabola of the free neutron, $Q^*=\sqrt{\frac{2m_n\omega^*}{\hbar^2}}=k_i$, with $\omega^*=E_i$, and thus the above integral can be reduced to:
\begin{equation}
2 \pi \frac{\hbar^2}{2m_nE_i}Q^*S^{down}(Q^*,+\omega^*)\int_{E_i-E_{ucn}}^{E_i} d\omega \int_{Q^*-k_{ucn}}^{Q^*+k_{ucn}} dQ  \nonumber
\end{equation}
\begin{equation}
=\frac{2 \pi \hbar^2}{2m_nE_i}Q^*S^{down}(Q^*,+\omega^*) E_{ucn} 2\Delta k_{ucn},
\label{eqn:UCNProd}
\end{equation}
given that S(Q, $\omega$) is a smooth function.
Next, to estimate the production of UCN that can be confined in the experimental apparatus, in which the Fermi-potential of the container material sets the maximum energy of UCN ($V_F=E_{ucn}$), an additional step is required to integrate the UCN energy from 0 to $V_F=\hbar^2k_{ucn}^2/2m$:
\begin{eqnarray}
\Delta k_{ucn} E_{ucn} &=&  \int_0^{k_{ucn}} dk_{ucn} \frac{\hbar^2k_{ucn}^2}{2m_n} \nonumber \\
&=& \frac{1}{3}\frac{\hbar^2k_{ucn}^3}{2m_n} = \frac{1}{3}k_{ucn}E_{ucn}. \nonumber
\end{eqnarray}

As a result, the cross-section of UCN production for neutrons with incident energy $E_i$ is 
\begin{equation}
\sigma(E_{i}\rightarrow (0-E_{ucn}))=\frac{1}{2}\frac{4 \pi \hbar^2}{2m_nE_i}k_iS^{down}(k_i,E_i) \frac{1}{3}k_{ucn}E_{ucn},
\end{equation} 
to produced storable UCN with $E_{ucn}$ from 0 to $E_{ucn}=V_F$.

\begin{figure}[t!]
\centering
\includegraphics[width=3.2 in]{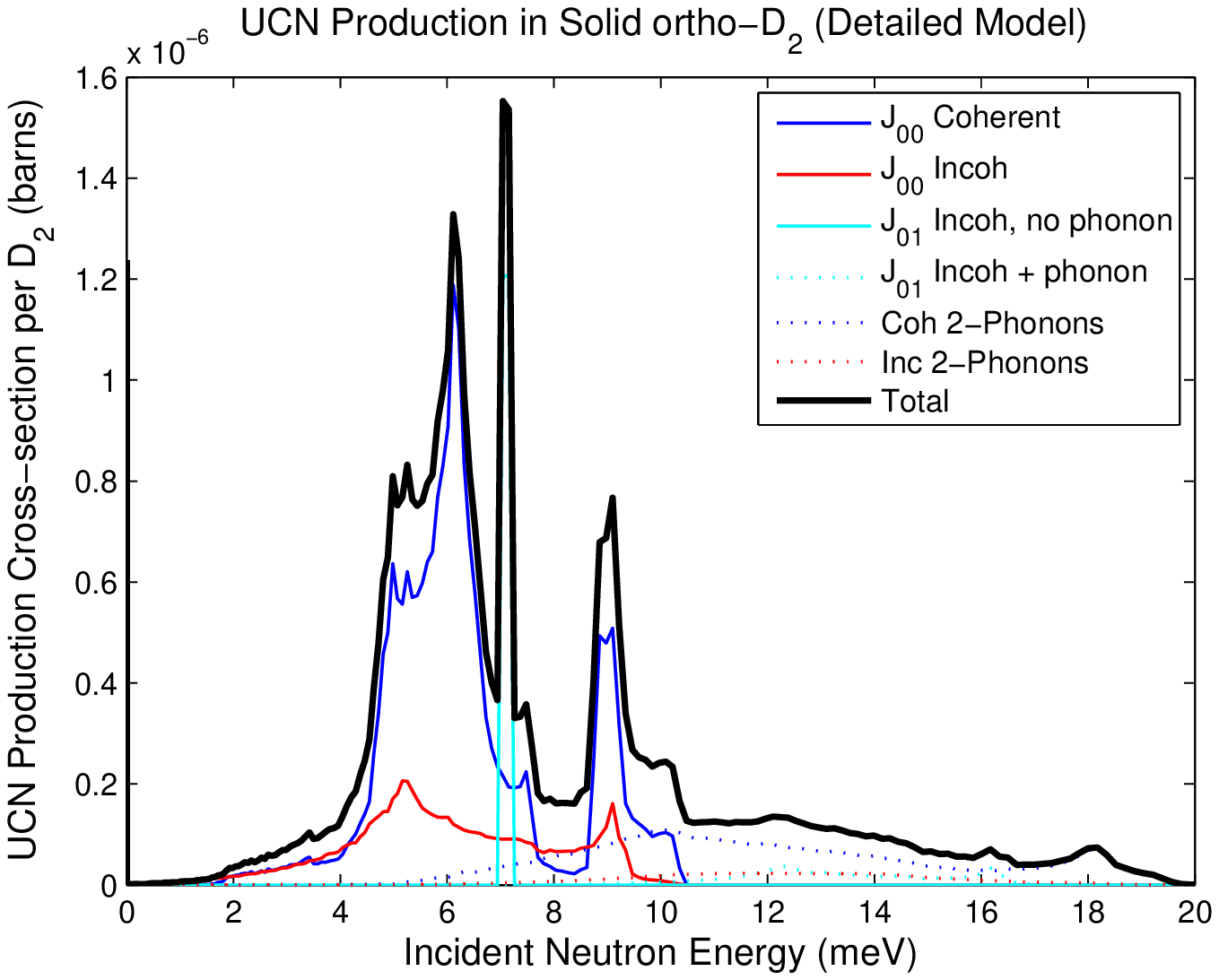}
\includegraphics[width=3.2 in]{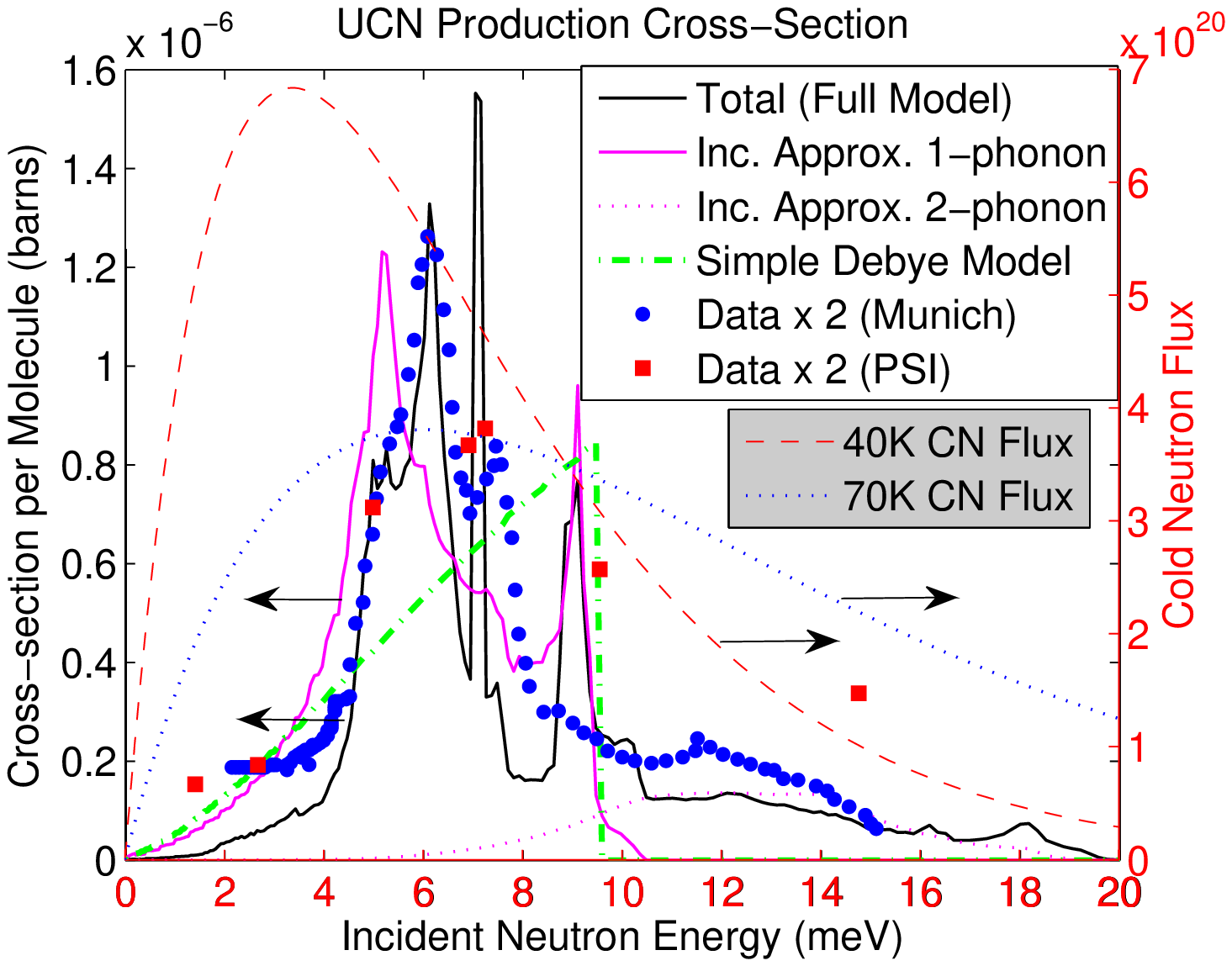}
\caption{\label{fig:UCNprod} {\bf Upper:} The UCN Production cross-section broken down into full different processes included in the full model, including coherent and incoherent excitations of 1 and 2 phonons, coupled to the two lowest rotational state transitions, i.e., J$_{00}$ and J$_{01}$. The J$_{01}$ process without phonon excitations also contributes to the UCN production. {\bf Lower:} The prediction using the full model is compared to that based on the IA and the simple Debye model, along with the experimental data. Energy spectra of cold neutron flux with Maxwell-Boltzmann distribution at 40~K and 70~K are plotted to illustrated the degree of overlap with the energy-dependent cross-sections.}
\end{figure}

Two sets of experimental data on UCN production are compared to the predictions of the full model.
First, note that the production cross-section reported in \cite{atchison2007} is normalized to each atom and their model does not include either the spin statistics or the molecular form factor of molecular D$_2$; the cross-section independently reported in \cite{Muller2008, Gutsmiedl2009} should be corrected by a factor of two to properly account for the range of integration on the allowed momentum transfer through $-k_{ucn}$ to $+k_{ucn}$.
To allow for direct comparisons, we multiply both data sets by a factor of two without any further corrections. 
Overall, the full model gives fair agreement with both the results of indirect UCN production extracted from the data of cold neutron scattering~\cite{Muller2008} and the results of direct UCN production \cite{atchison2007}(as shown in Fig.~\ref{fig:UCNprod}). Around the peak production at 6~meV, the full model gives a striking agreement with the experimental data, whereas the IA approach using a somewhat realistic density of states misses the peak by 1~meV. The simple Debye model fails to capture any details of energy dependence. 
The $J_{01}$ transition at 7.1~meV is a delta function in the model, but the finite energy resolution of the scattering instrument integrating around the transition energy reduces the amplitude.  
In spite of it, the data suggests that the transition energy is slightly higher than 7.1~meV, as used in the calculation. The second peak around 9~meV predicted by the full model is absent in the M\"{u}nich data \cite{Muller2008}, however, is present in the PSI data. The disappearance of the 9~meV peak could be a result of the preferred crystal orientation as discussed in Sec.\ref{sec:inel}. By restricting the scattering angle around the basal plane, the scattering intensity of the 9~meV peak can be adjusted to a smaller value (as shown in Fig.~\ref{fig:crystal}). 
For low scattering intensities at energy range below 4~meV and higher than 11~meV, the full model predicts production cross-sections smaller than the reported experimental data. This might simply due to the artifact of insufficient background subtraction in the experimental data. In particular, the subtraction of non-vanishing tails of the large elastic peaks is difficult to carry out. 
As for the high energy end, multiple scatterings couple to neutrons of energies higher than the Debye temperature, and thus enhance the scattering amplitudes. Neither of the data points is corrected for effects of multiple scattering.
The very different dimension of the target used in these two experiments also hints that the origin of the high energy excess might be multiple scattering in origin.

\begin{figure}[t!]
\includegraphics[width=3.0 in]{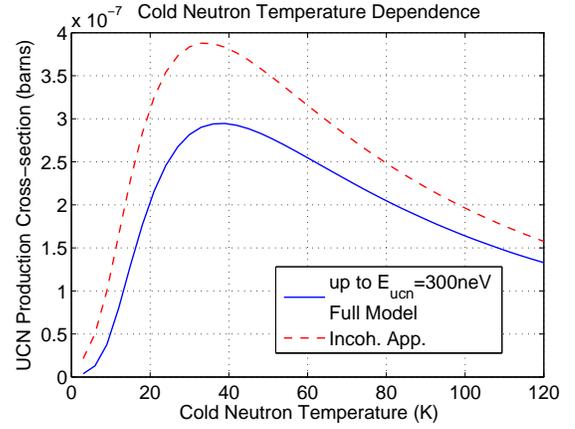}
\caption{The optimization of UCN Production by varying the incident cold neutron spectrum. The optimized cold neutron flux is a 33~K Maxwellian for IA treatment. The optimized cold neutron spectrum is a 40~K Maxwellian evaluated with the full model.}
\label{fig:UCNProd_CN}
\end{figure}

UCN production can be optimized by varying the energy spectrum of the incident cold neutrons.
Integrating the product of the energy-dependent cross-section for UCN production and the incident neutron flux over the energy spectrum yields a UCN production cross-section, 
\begin{equation}
\sigma(T_{cn})=\frac{1}{2}\int \frac{d\sigma(E_i\rightarrow E_{ucn})}{dE_i} \frac{E_i}{(k_{B}T_{cn})^2}e^{-E_i/k_BT_{cn}} dE_i ,
\end{equation} 
that depends on the temperature of incident cold neutrons.
Here we assume that the cold neutron flux has a thermalized energy spectrum with temperature $T_{cn}$. As shown in Fig.~\ref{fig:UCNProd_CN}, the UCN production is the greatest when coupled to a flux with a 40~K Maxwell-Boltzmann energy spectrum, whereas the optimized cold neutron spectrum predicted by the IA is 33~K with a cross-section 32\% larger than that predicted by the full model.
This difference comes from the fact that the IA approach over-estimates the contributions below 5~meV, where the free neutron parabola does not intersect with most phonon branches at low $Q$ (see Sec.~\ref{sec:inel}). Even though the differential cross-section for UCN production peaks at 6~meV, the overlap with a flux of cold neutrons with a 70~K spectrum does not result in more UCN production, because many neutrons in the Boltzmann distribution are spread out over the higher energy range resulting in a reduction of peaked flux. Using a colder flux of neutrons, a higher percentage of neutrons are directly under the production peak.   

\begin{figure}[t!]
\centering
\includegraphics[width=3.5 in]{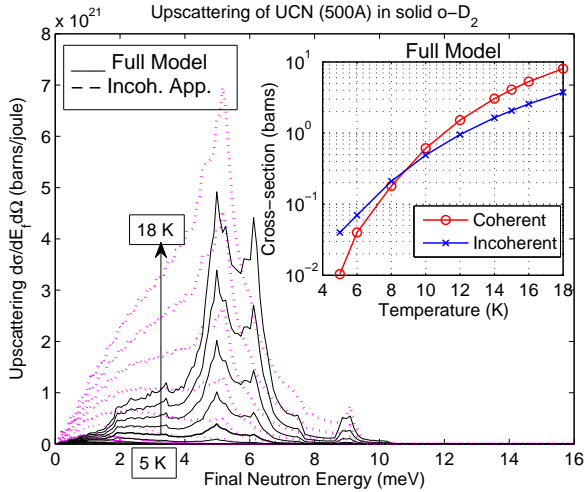}
\caption{\label{fig:UCNupEvolution}The evolution of the differential cross-section $d^2\sigma (E_{ucn}\rightarrow E_f)/dE_fd\Omega$ for UCN upscattering with increasing source temperature. Results of the full model (solid curves) and the IA predictions (magenta dashed curves) are plotted as a function of the final energy of upscattered neutrons. The inset plots the energy-integrated contributions of the coherent and incoherent process in the full model, as a function of the source temperature. }
\end{figure}

The achievable UCN density does not only depend on the production cross-section, but also scales with the lifetime of UCN, which sets the limit on the time duration during which the UCN density can accumulate without loss inside the source. Upscattering is a source of loss that needs to be controlled. 
Following the same approach for UCN production, the cross-section for UCN upscattering is evaluated by finding the $S(Q, -\omega)$ along the free neutron parabola as illustrated in the lower graph in Fig.~\ref{fig:SSDT}. 
Note that the upscattering amplitude is directly proportional to the occupation number of phonons, which obey the Bose-Einstein statistics. The occupation number peaks at low $\omega$ and falls off in magnitude as $\omega$ increases. 
The amplitude of UCN upscattering depends even more strongly on the physical presence of these phonons than that of the downscattering.
For small momentum transfers within the first Brillouin zone, the free neutrons do not intersect with any branches of acoustic phonon (see the lower graph in Fig.\ref{fig:SSDT}). Without direct coupling to the acoustic phonons, there is no coherent process that upscatters UCN though phonon annihilation. 
For solids at low temperatures, the non-zero upscattering amplitude comes from the remaining incoherent 1-phonon and multi-phonon processes (see Fig.~\ref{fig:UCNupEvolution}).
As the temperature increases, the contribution from coherent scattering increases with increasing thermal population of phonons beyond the first Brillouin zone (shown in the inset plot in Fig.~\ref{fig:UCNupEvolution}), and the coherent phonon annihilation take place. Fig.~\ref{fig:UCNupEvolution} presents the evolution of the differential cross-section of UCN upscattering with increasing temperature from 5~K to 18~K.

\begin{figure}
\centering
\includegraphics[width=3.2 in]{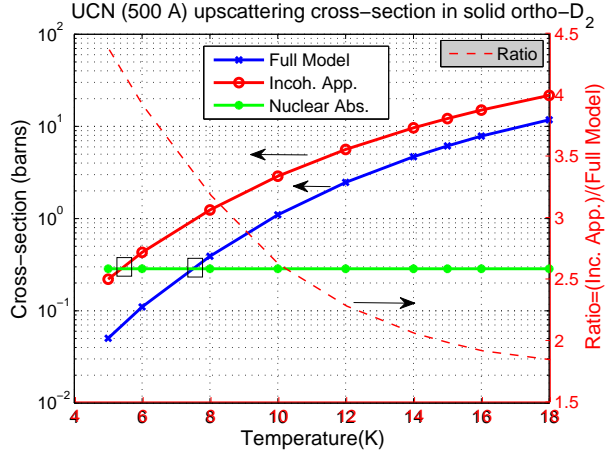}
\caption{\label{fig:UCNup}UCN upscattering cross-section (per D$_2$ molecule) in solid ortho-D$_2$ as a function of the source temperature. Both the full model (blue line with $\times$) and the IA predictions (red line with open circles) are plotted. The nuclear absorption (green line with solid circles) is independent of the source temperature. The intersection points between the upscattering cross-section and the nuclear absorption cross-section (square box) sets the saturation temperature of UCN yield. The ratio between the IA and full model prediction is plotted (dashed line) and presented along the y-axis on the right. }
\end{figure}

The total cross-section for UCN upscattering is calculated by integrating $S^{up}(Q,-\omega)$ (lower graph in Fig.\ref{fig:SSDT}) following: 
\begin{equation}
\sigma(E_{ucn}\rightarrow E_f)= \frac{1}{2} \int dE_f \int d\Omega \frac{k_f}{k_{ucn}}S^{up}(Q,-\omega).
\label{eqn:UCNup}
\end{equation}
Upon upscattering, UCN scatters into a narrow band of phase space with the final energy and momentum around the free neutron parabola (as illustrated in the lower graph in Fig.~\ref{fig:SSDT}). The integration simplifies into: 
\begin{align}
& \sigma(E_{ucn}\rightarrow \mbox{all possible }E_f) = \nonumber \\
& \frac{1}{2}\int d\phi \int_{E_{ucn}}^{\omega^*+E_{ucn}} d\omega \int_{Q^*-k_{ucn}}^{Q^*+k_{ucn}} dQ Q^* \frac{1}{k^2_i}S^{up}(Q^*,-\omega^*) \nonumber \\
&\hspace{0.6 in}= \frac{2\pi}{2} \int_0^{\infty} dE_f Q^*\frac{1}{k^2_{ucn}}S^{up}(Q^*,-\omega^*)2k_{ucn} \nonumber \\
&\hspace{0.6 in}= \frac{4\pi}{2} \frac{1}{k_{ucn}}\int_0^{\infty} dE_f k_fS^{up}(k_f,-E_f).  
\end{align}
In Fig.~\ref{fig:UCNup} plots upscattering cross-section for UCN with a wavelength of 500 $\buildrel _\circ \over {\mathrm{A}}$. 
For UCN with a different wavelength, the upscattering is rescaled linearly with its initial wavelength. 
The full model predicts a UCN upscattering cross-section 2$\sim$4 times smaller than that estimated by IA (see Fig.~\ref{fig:UCNup}) depending on the temperature.  
The excess in the IA prediction is due to the use of the density of states that over-estimates the number of modes in small energy transfers within the first Brillouin zone (see details shown in Fig.~\ref{fig:UCNupEvolution}).
This correction is by no means small. 
With a reduced upscattering cross-section, UCN can live longer inside the source at higher temperatures. 

The ultimate limit on the UCN lifetime in solid D$_2$ comes from the small, yet non-vanishing neutron absorption of each deuteron nuclide. In a source where the escape time of UCN is comparable to (or larger than) the absorption time, the UCN yield saturates when the upscattering loss is reduced to the same level as the nuclear absorption loss. Reduction of the upscattering by further cooling would not increase the UCN yield of UCN by more than a factor of two. 
In practice, the gain is smaller than two because of the additional sources of loss, such as the hydrogen contamination and the para-D$_2$ contamination~\cite{liu2000,morris2002}.
The condition for saturation in UCN output sets the practical operational temperature of the source, and determines the requirements on cryogenics engineering. 
The direct consequence of reduced upscattering cross-section in solid ortho-D$_2$ is that the predicted saturation temperature is $\sim$ 7.5~K, which is 2 degrees higher than that predicted previously using the IA. 
The higher operating temperature makes the implementation of cryogenic source somewhat less challenging as most refrigerator systems have higher cooling power at elevated temperatures.

For a practical sample of converted D$_2$ gas containing a residual 3\% of para-D$_2$, the total loss cross-section is increased by 1~b, because the temperature-independent upscattering cross-section of pure para-D$_2$ is 31~b~\cite{liu2000}. In this case, the full model predicts the saturation temperature for UCN production to be 11~K, and the IA model predicts it to be 8~K. 
Using un-converted normal D$_2$ gas with 33\% para-D$_2$, the loss cross-section is about 10~b. According to the upscattering cross-sections presented in Fig.~\ref{fig:UCNup}, the UCN yield would reach saturation at temperatures as high as 18~K, and no superthermal gain would be observed by cooling the UCN converter.   

\begin{figure}[t!]
\centering
\includegraphics[width=3.5 in]{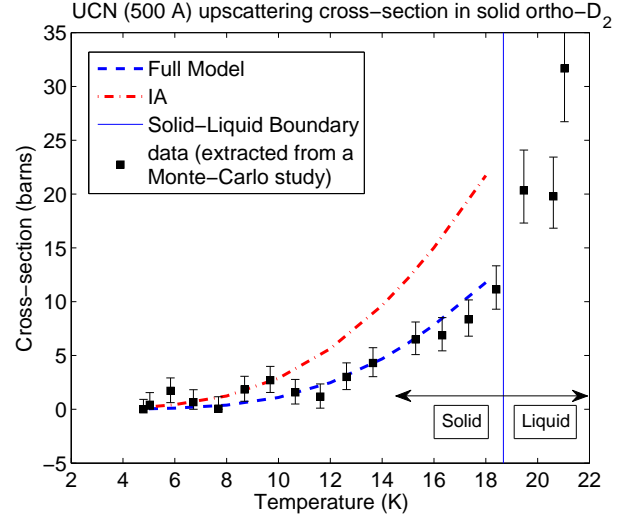}
\caption{\label{fig:UpExp} Upscattering cross-sections (for 500 \mbox{\AA} UCN) measured in a UCN experiment~\protect\cite{Lavelle2010}, compared to predictions from the full model and the IA. The vertical line indicates the triple point of solid D$_2$ at temperature of 18.678~K, separating the solid and liquid phases.}
\end{figure}

Finally, the updated upscattering cross-section predicted by the full model of $S(Q, \omega)$ for o-D$_2$ is applied to understand the UCN production data from our recent measurements using solid D$_2$ \cite{Lavelle2010}. The Monte-Carlo simulation using the upscattering cross-sections calculated using the IA approach fails to reproduce the temperature dependence measured in the experiment. 
Details of the Monte-Carlo simulation can be found in \cite{Lavelle2010}.
We also investigated other effects that lead to increased elastic scattering inside the source, however, we find that the higher saturation temperature can only be explained by reduced loss. The coherent scattering leading to smaller upscattering cross-sections predicted using our full model is a very likely candidate to provide the required modification.

We then extract upscattering cross-sections using the same set of experimental data reported in \cite{Lavelle2010}.
The upscattering cross-sections of solid D$_2$ are derived from another Monte-Carlo study, in which the upscattering cross-section is varied until the simulated result on UCN yield best fits the experimental data for each different temperature. Results of this study are shown in Fig.~\ref{fig:UpExp}, compared to theoretical predictions based on the full model and the IA. 
The Monte-Carlo code includes the full range of UCN energy spectrum from 0 up to 1~$\mu$eV. To facilitate direct comparison with previous calculations, the extracted upscattering cross-section are for UCN with a wavelength of 500~\mbox{\AA}. 
Below the triple point, the upscattring cross-sections agree quite well with the full model prediction. Above the triple point, the mechanism for UCN upscattering (in the liquid phase) is significantly different from that in the solid, and none of the models discussed in this paper apply. 
Data points of small upscattering cross-sections fluctuates at low temperatures due to low statistics of UCN signal in general.
At temperatures higher than 10~K, where the UCN upscattering cross-section is larger than 1 b, the incoherent model is excluded at the 2$\sigma$ level and higher.

\begin{figure}[t!]
\centering
\includegraphics[width=3.5 in]{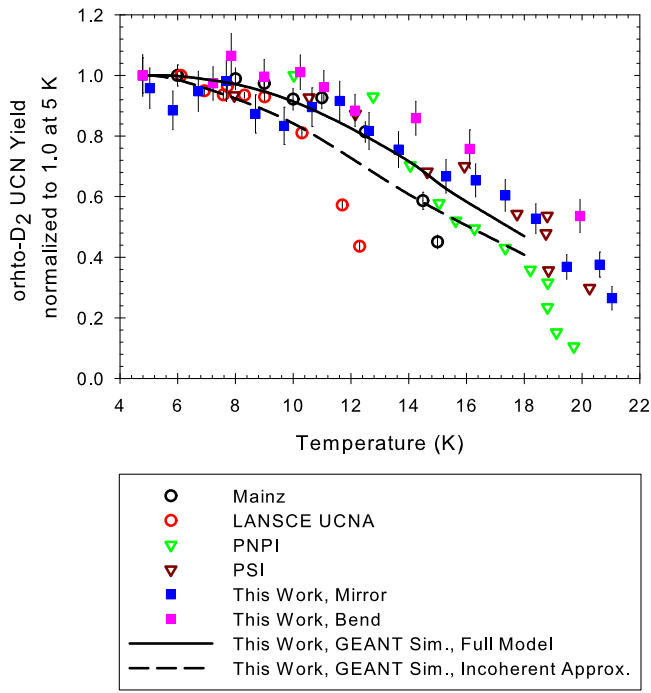}
\caption{\label{fig:Exp_all} The temperature dependence of our FP12 experiment data reported in \protect\cite{Lavelle2010} is compared to similar experiments (Mainz\protect\cite{lauer_thesis}, LANSCE UCNA\protect\cite{morris2002}, PNPI\protect\cite{serebrov2000a}, PSI\protect\cite{atchison2005a}).  All data are scaled to unity at 5 K. GEANT4 simulations using upscattering cross-sections calculated by the full model (solid curve) is compared to that using the IA (dashed curve).}
\end{figure}

We also surveyed experimental results on UCN production using solid deuterium reported independently by various research groups. Data sets are plotted in Fig.~\ref{fig:Exp_all} together with the results of simulations.
While the results of simulation using the updated upscattering cross-section agree quite well with our data and the data set measured at PSI, the data sets from LANL and Mainz groups show a much steeper temperature dependence for temperatures higher than 10~K. The difference comes from the different source configurations. In these two experiments, solid D$_2$ was condensed from vapor at the end of the UCN guide, which was cooled below the solidification temperature of D$_2$. The source was designed to reduce the transmission loss by eliminating the vacuum window that is typically installed to contain the volatile D$_2$. The windowless source worked quite well at low temperatures, however, for temperatures higher than 10~K, the entire UCN guide was filled with D$_2$ gas at the saturated vapor pressure. The large upscattering cross-section from the D$_2$ vapor, together with a large volume where the vapor permeated, resulted in a steeper temperature dependence than the simple prediction where UCN upscatters through absorption of phonon in the confined region of the source.
Both PSI and our experiment used D$_2$ contained in a target cell, in which the vapor region is considerably smaller.
None of the data sets show a low saturation temperature, as predicted by the IA. Instead, the temperature dependence of experimental data points agrees better with results of the simulation using upscattering cross-sections predicted by the full model.  




\section{Conclusion}
In this paper, we have presented calculations of the neutron cross-sections in polycrystalline ortho-D$_2$. All known physics on spin statistics, rotational excitations, elastic Bragg scattering, coherent inelastic, and incoherent inelastic scattering have been implemented in our full model for the first time. The presented work addresses the shortcomings of the widely used IA in estimating neutron cross-sections at energies at thermal and cold regime. For applications on modern neutron sources for UCN production using solid D$_2$ and other materials with large coherent scattering lengths, such as helium, oxygen and nitrogen, the interference between scattered neutron wavefunctions across all lattice sites is important and should not be ignored. For the case of solid ortho-D$_2$, we have shown that UCN cross-sections are significantly altered under the new treatment. In particular, the UCN upscattering cross-section is a factor of 2$\sim$4 smaller than previous predictions throughout the temperature range of interest to UCN production.
The same modifications will impact designs of sources (beyond UCN sources) that aim to produce long wavelength neutrons using materials other than hydrogen. The Monte-Carlo algorithm developed in this work can be applied to all materials with significant coherent scattering components, provided that the energy dispersion of the quasi-particles is known.


\begin{acknowledgements}
The work was supported by NSF grants 0457219, 0758018.

\end{acknowledgements}

\bibliography{D2paper}

\begin{thebibliography}{41}
\expandafter\ifx\csname natexlab\endcsname\relax\def\natexlab#1{#1}\fi
\expandafter\ifx\csname bibnamefont\endcsname\relax
  \def\bibnamefont#1{#1}\fi
\expandafter\ifx\csname bibfnamefont\endcsname\relax
  \def\bibfnamefont#1{#1}\fi
\expandafter\ifx\csname citenamefont\endcsname\relax
  \def\citenamefont#1{#1}\fi
\expandafter\ifx\csname url\endcsname\relax
  \def\url#1{\texttt{#1}}\fi
\expandafter\ifx\csname urlprefix\endcsname\relax\def\urlprefix{URL }\fi
\providecommand{\bibinfo}[2]{#2}
\providecommand{\eprint}[2][]{\url{#2}}

\bibitem[{\citenamefont{Glasstone and Edlund}(1952)}]{Glasstone52}
\bibinfo{author}{\bibfnamefont{S.}~\bibnamefont{Glasstone}} \bibnamefont{and}
  \bibinfo{author}{\bibfnamefont{M.}~\bibnamefont{Edlund}},
  \emph{\bibinfo{title}{The Elements of NUCLEAR REACTOR THEORY}}
  (\bibinfo{publisher}{D. Van Nostrand Company, Inc.}, \bibinfo{address}{New
  York}, \bibinfo{year}{1952}).

\bibitem[{\citenamefont{Labratory}(2008)}]{mcnp5}
\bibinfo{author}{\bibfnamefont{L.~A.~N.} \bibnamefont{Labratory}},
  \emph{\bibinfo{title}{Mcnp5 monte carlo code, ccc-730}}
  (\bibinfo{year}{2008}).

\bibitem[{\citenamefont{MacFarlane}(1994)}]{MacFarlane94}
\bibinfo{author}{\bibfnamefont{R.}~\bibnamefont{MacFarlane}},
  \emph{\bibinfo{title}{New thermal neutron scattering files for endf/b-vi
  release 2}} (\bibinfo{year}{1994}).

\bibitem[{\citenamefont{Young and Koppel}(1964)}]{Young64}
\bibinfo{author}{\bibfnamefont{J.~A.} \bibnamefont{Young}} \bibnamefont{and}
  \bibinfo{author}{\bibfnamefont{J.~U.} \bibnamefont{Koppel}},
  \bibinfo{journal}{Physical Review a-General Physics}
  \textbf{\bibinfo{volume}{135}}, \bibinfo{pages}{A603} (\bibinfo{year}{1964}).

\bibitem[{\citenamefont{Schwinger}(1940)}]{Schwinger37}
\bibinfo{author}{\bibfnamefont{J.}~\bibnamefont{Schwinger}},
  \bibinfo{journal}{Physical Review} \textbf{\bibinfo{volume}{58}},
  \bibinfo{pages}{1004} (\bibinfo{year}{1940}).

\bibitem[{\citenamefont{Hamermesh and Schwinger}(1946)}]{Hamermesh46}
\bibinfo{author}{\bibfnamefont{M.}~\bibnamefont{Hamermesh}} \bibnamefont{and}
  \bibinfo{author}{\bibfnamefont{J.}~\bibnamefont{Schwinger}},
  \bibinfo{journal}{Physical Review} \textbf{\bibinfo{volume}{69}},
  \bibinfo{pages}{145} (\bibinfo{year}{1946}).

\bibitem[{njo(1999)}]{njoy99}
\emph{\bibinfo{title}{Njoy -- nuclear data processing system}}
  (\bibinfo{year}{1999}).

\bibitem[{\citenamefont{Liu et~al.}(2000)\citenamefont{Liu, Young, and
  Lamoreaux}}]{liu2000}
\bibinfo{author}{\bibfnamefont{C.~Y.} \bibnamefont{Liu}},
  \bibinfo{author}{\bibfnamefont{A.~R.} \bibnamefont{Young}}, \bibnamefont{and}
  \bibinfo{author}{\bibfnamefont{S.~K.} \bibnamefont{Lamoreaux}},
  \bibinfo{journal}{Physical Review B} \textbf{\bibinfo{volume}{62}},
  \bibinfo{pages}{R3581} (\bibinfo{year}{2000}).

\bibitem[{\citenamefont{Atchison et~al.}(2007)\citenamefont{Atchison, Blau,
  Bodek, van~den Brandt, Brys, Daum, Fierlinger, Frei, Geltenbort, Hautle
  et~al.}}]{atchison2007}
\bibinfo{author}{\bibfnamefont{F.}~\bibnamefont{Atchison}},
  \bibinfo{author}{\bibfnamefont{B.}~\bibnamefont{Blau}},
  \bibinfo{author}{\bibfnamefont{K.}~\bibnamefont{Bodek}},
  \bibinfo{author}{\bibfnamefont{B.}~\bibnamefont{van~den Brandt}},
  \bibinfo{author}{\bibfnamefont{T.}~\bibnamefont{Brys}},
  \bibinfo{author}{\bibfnamefont{M.}~\bibnamefont{Daum}},
  \bibinfo{author}{\bibfnamefont{P.}~\bibnamefont{Fierlinger}},
  \bibinfo{author}{\bibfnamefont{A.}~\bibnamefont{Frei}},
  \bibinfo{author}{\bibfnamefont{P.}~\bibnamefont{Geltenbort}},
  \bibinfo{author}{\bibfnamefont{P.}~\bibnamefont{Hautle}},
  \bibnamefont{et~al.}, \bibinfo{journal}{Physical Review Letters}
  \textbf{\bibinfo{volume}{99}}, \bibinfo{pages}{262502}
  (\bibinfo{year}{2007}).

\bibitem[{\citenamefont{Granada}(2009)}]{Granada09}
\bibinfo{author}{\bibfnamefont{J.~R.} \bibnamefont{Granada}},
  \bibinfo{journal}{EPL} \textbf{\bibinfo{volume}{86}}, \bibinfo{pages}{66007}
  (\bibinfo{year}{2009}).

\bibitem[{\citenamefont{Egelstaff et~al.}(1967)\citenamefont{Egelstaff,
  Haywood, and Webb}}]{Egelstaff67}
\bibinfo{author}{\bibfnamefont{P.~A.} \bibnamefont{Egelstaff}},
  \bibinfo{author}{\bibfnamefont{B.~C.} \bibnamefont{Haywood}},
  \bibnamefont{and} \bibinfo{author}{\bibfnamefont{F.~J.} \bibnamefont{Webb}},
  \bibinfo{journal}{Proceedings of the Physical Society of London}
  \textbf{\bibinfo{volume}{90}}, \bibinfo{pages}{681} (\bibinfo{year}{1967}).

\bibitem[{\citenamefont{Elliott and Hartmann}(1967)}]{Elliott67}
\bibinfo{author}{\bibfnamefont{R.~J.} \bibnamefont{Elliott}} \bibnamefont{and}
  \bibinfo{author}{\bibfnamefont{W.~M.} \bibnamefont{Hartmann}},
  \bibinfo{journal}{Proceedings of the Physical Society of London}
  \textbf{\bibinfo{volume}{90}}, \bibinfo{pages}{671} (\bibinfo{year}{1967}).

\bibitem[{\citenamefont{Diehl and Biem}(1975)}]{Diehl75}
\bibinfo{author}{\bibfnamefont{H.~W.} \bibnamefont{Diehl}} \bibnamefont{and}
  \bibinfo{author}{\bibfnamefont{W.}~\bibnamefont{Biem}},
  \bibinfo{journal}{Zeitschrift Fur Physik B-Condensed Matter}
  \textbf{\bibinfo{volume}{20}}, \bibinfo{pages}{137} (\bibinfo{year}{1975}).

\bibitem[{\citenamefont{Danchuk et~al.}(2004)\citenamefont{Danchuk, Galtsov,
  Strzhemechny, and Prokhvatilov}}]{Danchuk04}
\bibinfo{author}{\bibfnamefont{V.~V.} \bibnamefont{Danchuk}},
  \bibinfo{author}{\bibfnamefont{N.~N.} \bibnamefont{Galtsov}},
  \bibinfo{author}{\bibfnamefont{M.~A.} \bibnamefont{Strzhemechny}},
  \bibnamefont{and} \bibinfo{author}{\bibfnamefont{A.~I.}
  \bibnamefont{Prokhvatilov}}, \bibinfo{journal}{Low Temperature Physics}
  \textbf{\bibinfo{volume}{30}}, \bibinfo{pages}{118} (\bibinfo{year}{2004}).

\bibitem[{\citenamefont{Squires and Stewart}(1955)}]{Squire55}
\bibinfo{author}{\bibfnamefont{G.~L.} \bibnamefont{Squires}} \bibnamefont{and}
  \bibinfo{author}{\bibfnamefont{A.~T.} \bibnamefont{Stewart}},
  \bibinfo{journal}{Proceedings of the Royal Society of London Series
  a-Mathematical and Physical Sciences} \textbf{\bibinfo{volume}{230}},
  \bibinfo{pages}{19} (\bibinfo{year}{1955}).

\bibitem[{\citenamefont{Schott}(1970)}]{Schott70}
\bibinfo{author}{\bibfnamefont{W.}~\bibnamefont{Schott}},
  \bibinfo{journal}{Zeitschrift Fur Physik} \textbf{\bibinfo{volume}{231}},
  \bibinfo{pages}{243} (\bibinfo{year}{1970}).

\bibitem[{\citenamefont{Nielsen and Møller}(1971)}]{Nielsen71}
\bibinfo{author}{\bibfnamefont{M.}~\bibnamefont{Nielsen}} \bibnamefont{and}
  \bibinfo{author}{\bibfnamefont{H.~B.} \bibnamefont{Møller}},
  \bibinfo{journal}{Physical Review B} \textbf{\bibinfo{volume}{3}},
  \bibinfo{pages}{4383} (\bibinfo{year}{1971}).

\bibitem[{\citenamefont{Nielsen}(1973)}]{Nielsen73}
\bibinfo{author}{\bibfnamefont{M.}~\bibnamefont{Nielsen}},
  \bibinfo{journal}{Physical Review B} \textbf{\bibinfo{volume}{7}},
  \bibinfo{pages}{1626} (\bibinfo{year}{1973}).

\bibitem[{\citenamefont{Schmidt et~al.}(1984)\citenamefont{Schmidt, Nielsen,
  and Daniels}}]{Schmidt84}
\bibinfo{author}{\bibfnamefont{J.~W.} \bibnamefont{Schmidt}},
  \bibinfo{author}{\bibfnamefont{M.}~\bibnamefont{Nielsen}}, \bibnamefont{and}
  \bibinfo{author}{\bibfnamefont{W.~B.} \bibnamefont{Daniels}},
  \bibinfo{journal}{Physical Review B} \textbf{\bibinfo{volume}{30}},
  \bibinfo{pages}{6308} (\bibinfo{year}{1984}).

\bibitem[{\citenamefont{Dewames et~al.}(1965)\citenamefont{Dewames, Wolfram,
  and Lehman}}]{Dewames65}
\bibinfo{author}{\bibfnamefont{R.~E.} \bibnamefont{Dewames}},
  \bibinfo{author}{\bibfnamefont{T.}~\bibnamefont{Wolfram}}, \bibnamefont{and}
  \bibinfo{author}{\bibfnamefont{G.~W.} \bibnamefont{Lehman}},
  \bibinfo{journal}{Physical Review} \textbf{\bibinfo{volume}{138}},
  \bibinfo{pages}{A717} (\bibinfo{year}{1965}).

\bibitem[{\citenamefont{Lehman et~al.}(1962)\citenamefont{Lehman, Dewames, and
  Wolfram}}]{Lehman62}
\bibinfo{author}{\bibfnamefont{G.~W.} \bibnamefont{Lehman}},
  \bibinfo{author}{\bibfnamefont{R.~E.} \bibnamefont{Dewames}},
  \bibnamefont{and} \bibinfo{author}{\bibfnamefont{T.}~\bibnamefont{Wolfram}},
  \bibinfo{journal}{Physical Review} \textbf{\bibinfo{volume}{128}},
  \bibinfo{pages}{1593} (\bibinfo{year}{1962}).

\bibitem[{\citenamefont{Collins}(1962)}]{Collins62}
\bibinfo{author}{\bibfnamefont{M.~F.} \bibnamefont{Collins}},
  \bibinfo{journal}{Proceedings of the Physical Society of London}
  \textbf{\bibinfo{volume}{80}}, \bibinfo{pages}{362} (\bibinfo{year}{1962}).

\bibitem[{\citenamefont{Houmann}(1970)}]{Houmann70}
\bibinfo{author}{\bibfnamefont{J.~C.~G.} \bibnamefont{Houmann}},
  \bibinfo{journal}{Physical Review B} \textbf{\bibinfo{volume}{1}},
  \bibinfo{pages}{3943} (\bibinfo{year}{1970}).

\bibitem[{\citenamefont{Vaks and Khromov}(2008)}]{Vaks08}
\bibinfo{author}{\bibfnamefont{V.~G.} \bibnamefont{Vaks}} \bibnamefont{and}
  \bibinfo{author}{\bibfnamefont{K.~Y.} \bibnamefont{Khromov}},
  \bibinfo{journal}{Journal of Experimental and Theoretical Physics}
  \textbf{\bibinfo{volume}{106}}, \bibinfo{pages}{495} (\bibinfo{year}{2008}).

\bibitem[{\citenamefont{Lovesey}(1984)}]{loveseyBook}
\bibinfo{author}{\bibfnamefont{S.}~\bibnamefont{Lovesey}},
  \emph{\bibinfo{title}{Theory of Neutron Scattering from Condensed Matter}},
  vol.~\bibinfo{volume}{1} (\bibinfo{publisher}{Oxford University Press},
  \bibinfo{address}{Oxford}, \bibinfo{year}{1984}).

\bibitem[{\citenamefont{Colognesi et~al.}(2009)\citenamefont{Colognesi,
  Formisano, Ramirez-Cuesta, and Ulivi}}]{Colognesi2009}
\bibinfo{author}{\bibfnamefont{D.}~\bibnamefont{Colognesi}},
  \bibinfo{author}{\bibfnamefont{F.}~\bibnamefont{Formisano}},
  \bibinfo{author}{\bibfnamefont{A.~J.} \bibnamefont{Ramirez-Cuesta}},
  \bibnamefont{and} \bibinfo{author}{\bibfnamefont{L.}~\bibnamefont{Ulivi}},
  \bibinfo{journal}{Physical Review B} \textbf{\bibinfo{volume}{79}},
  (\bibinfo{year}{2009}).

\bibitem[{\citenamefont{Egelstaff and Pease}(1954)}]{Egelstaff54}
\bibinfo{author}{\bibfnamefont{P.~A.} \bibnamefont{Egelstaff}}
  \bibnamefont{and} \bibinfo{author}{\bibfnamefont{R.~S.} \bibnamefont{Pease}},
  \bibinfo{journal}{Journal of Scientific Instruments}
  \textbf{\bibinfo{volume}{31}}, \bibinfo{pages}{207} (\bibinfo{year}{1954}).

\bibitem[{\citenamefont{Bostanjo.O and Kleinsch.R}(1967)}]{Bostanjo67}
\bibinfo{author}{\bibnamefont{Bostanjo.O}} \bibnamefont{and}
  \bibinfo{author}{\bibnamefont{Kleinsch.R}}, \bibinfo{journal}{Journal of
  Chemical Physics} \textbf{\bibinfo{volume}{46}}, \bibinfo{pages}{2004}
  (\bibinfo{year}{1967}).

\bibitem[{\citenamefont{Collins et~al.}(1996)\citenamefont{Collins, Unites,
  Mapoles, and Bernat}}]{Collins96}
\bibinfo{author}{\bibfnamefont{G.~W.} \bibnamefont{Collins}},
  \bibinfo{author}{\bibfnamefont{W.~G.} \bibnamefont{Unites}},
  \bibinfo{author}{\bibfnamefont{E.~R.} \bibnamefont{Mapoles}},
  \bibnamefont{and} \bibinfo{author}{\bibfnamefont{T.~P.}
  \bibnamefont{Bernat}}, \bibinfo{journal}{Physical Review B}
  \textbf{\bibinfo{volume}{53}}, \bibinfo{pages}{102} (\bibinfo{year}{1996}).

\bibitem[{\citenamefont{Stein et~al.}(1972)\citenamefont{Stein, Stiller, and
  Stockmey.R}}]{Stein72}
\bibinfo{author}{\bibfnamefont{H.}~\bibnamefont{Stein}},
  \bibinfo{author}{\bibfnamefont{H.}~\bibnamefont{Stiller}}, \bibnamefont{and}
  \bibinfo{author}{\bibnamefont{Stockmey.R}}, \bibinfo{journal}{Journal of
  Chemical Physics} \textbf{\bibinfo{volume}{57}}, \bibinfo{pages}{1726}
  (\bibinfo{year}{1972}).

\bibitem[{\citenamefont{Seiffert}(1970)}]{seiffert1970}
\bibinfo{author}{\bibfnamefont{W.-D.} \bibnamefont{Seiffert}},
  \bibinfo{type}{Tech. Rep.}, \bibinfo{institution}{Euroa\"{a}ische
  atomgemeinschaft euratom} (\bibinfo{year}{1970}).

\bibitem[{\citenamefont{Lavelle et~al.}(2010)\citenamefont{Lavelle, Fox, Manus,
  McChesney, Salvat, Shin, Makela, Morris, Saunders, Couture
  et~al.}}]{Lavelle2010}
\bibinfo{author}{\bibfnamefont{C.}~\bibnamefont{Lavelle}},
  \bibinfo{author}{\bibfnamefont{W.}~\bibnamefont{Fox}},
  \bibinfo{author}{\bibfnamefont{G.}~\bibnamefont{Manus}},
  \bibinfo{author}{\bibfnamefont{P.}~\bibnamefont{McChesney}},
  \bibinfo{author}{\bibfnamefont{D.}~\bibnamefont{Salvat}},
  \bibinfo{author}{\bibfnamefont{Y.}~\bibnamefont{Shin}},
  \bibinfo{author}{\bibfnamefont{M.}~\bibnamefont{Makela}},
  \bibinfo{author}{\bibfnamefont{C.}~\bibnamefont{Morris}},
  \bibinfo{author}{\bibfnamefont{A.}~\bibnamefont{Saunders}},
  \bibinfo{author}{\bibfnamefont{A.}~\bibnamefont{Couture}},
  \bibnamefont{et~al.}, \emph{\bibinfo{title}{Ultracold neutron production in a
  pulsed neutron beam line}} (\bibinfo{year}{2010}).

\bibitem[{\citenamefont{Liu and Young}(2004)}]{Liu2004}
\bibinfo{author}{\bibfnamefont{C.~Y.} \bibnamefont{Liu}} \bibnamefont{and}
  \bibinfo{author}{\bibfnamefont{A.}~\bibnamefont{Young}},
  \bibinfo{journal}{arxiv:nucl-th/0406004}  (\bibinfo{year}{2004}),
  \bibinfo{note}{submitted to Phys. Rev. B}.

\bibitem[{\citenamefont{Yu et~al.}(1986)\citenamefont{Yu, Malik, and
  Golub}}]{Yu86}
\bibinfo{author}{\bibfnamefont{Z.~C.} \bibnamefont{Yu}},
  \bibinfo{author}{\bibfnamefont{S.~S.} \bibnamefont{Malik}}, \bibnamefont{and}
  \bibinfo{author}{\bibfnamefont{R.}~\bibnamefont{Golub}},
  \bibinfo{journal}{Zeitschrift Fur Physik B-Condensed Matter}
  \textbf{\bibinfo{volume}{62}}, \bibinfo{pages}{137} (\bibinfo{year}{1986}).

\bibitem[{\citenamefont{Raubenheimer and Gilat}(1967)}]{Rabubenheimer1967}
\bibinfo{author}{\bibfnamefont{L.~J.} \bibnamefont{Raubenheimer}}
  \bibnamefont{and} \bibinfo{author}{\bibfnamefont{G.}~\bibnamefont{Gilat}},
  \bibinfo{journal}{Physical Review} \textbf{\bibinfo{volume}{157}},
  \bibinfo{pages}{586} (\bibinfo{year}{1967}).

\bibitem[{\citenamefont{Gutsmiedl et~al.}(2009)\citenamefont{Gutsmiedl, Frei,
  Müller, Paul, Urban, Schober, Morkel, and Unruh}}]{Gutsmiedl2009}
\bibinfo{author}{\bibfnamefont{E.}~\bibnamefont{Gutsmiedl}},
  \bibinfo{author}{\bibfnamefont{A.}~\bibnamefont{Frei}},
  \bibinfo{author}{\bibfnamefont{A.~R.} \bibnamefont{Müller}},
  \bibinfo{author}{\bibfnamefont{S.}~\bibnamefont{Paul}},
  \bibinfo{author}{\bibfnamefont{M.}~\bibnamefont{Urban}},
  \bibinfo{author}{\bibfnamefont{H.}~\bibnamefont{Schober}},
  \bibinfo{author}{\bibfnamefont{C.}~\bibnamefont{Morkel}}, \bibnamefont{and}
  \bibinfo{author}{\bibfnamefont{T.}~\bibnamefont{Unruh}},
  \bibinfo{journal}{Nuclear Instruments and Methods in Physics Research Section
  A: Accelerators, Spectrometers, Detectors and Associated Equipment}
  \textbf{\bibinfo{volume}{611}}, \bibinfo{pages}{256} (\bibinfo{year}{2009}).

\bibitem[{\citenamefont{Muller}(2008)}]{Muller2008}
\bibinfo{author}{\bibfnamefont{A.}~\bibnamefont{Muller}}, Ph.D. thesis
  (\bibinfo{year}{2008}).

\bibitem[{\citenamefont{Thorsten}(2010)}]{lauer_thesis}
\bibinfo{author}{\bibfnamefont{L.}~\bibnamefont{Thorsten}}, Ph.D. thesis
  (\bibinfo{year}{2010}).

\bibitem[{\citenamefont{Morris et~al.}(2002)\citenamefont{Morris, Anaya,
  Bowles, Filippone, Geltenbort, Hill, Hino, Hoedl, Hogan, Ito
  et~al.}}]{morris2002}
\bibinfo{author}{\bibfnamefont{C.~L.} \bibnamefont{Morris}},
  \bibinfo{author}{\bibfnamefont{J.~M.} \bibnamefont{Anaya}},
  \bibinfo{author}{\bibfnamefont{T.~J.} \bibnamefont{Bowles}},
  \bibinfo{author}{\bibfnamefont{B.~W.} \bibnamefont{Filippone}},
  \bibinfo{author}{\bibfnamefont{P.}~\bibnamefont{Geltenbort}},
  \bibinfo{author}{\bibfnamefont{R.~E.} \bibnamefont{Hill}},
  \bibinfo{author}{\bibfnamefont{M.}~\bibnamefont{Hino}},
  \bibinfo{author}{\bibfnamefont{S.}~\bibnamefont{Hoedl}},
  \bibinfo{author}{\bibfnamefont{G.~E.} \bibnamefont{Hogan}},
  \bibinfo{author}{\bibfnamefont{T.~M.} \bibnamefont{Ito}},
  \bibnamefont{et~al.}, \bibinfo{journal}{Physical Review Letters}
  \textbf{\bibinfo{volume}{89}}, \bibinfo{pages}{272501}
  (\bibinfo{year}{2002}).

\bibitem[{\citenamefont{Serebrov et~al.}(2000)\citenamefont{Serebrov,
  Mityukhlyaev, Zakharov, Kharitonov, Shustov, Kuz'minov, Lasakov, Tal'daev,
  Aldushchenkov, Varlamov et~al.}}]{serebrov2000a}
\bibinfo{author}{\bibfnamefont{A.}~\bibnamefont{Serebrov}},
  \bibinfo{author}{\bibfnamefont{V.}~\bibnamefont{Mityukhlyaev}},
  \bibinfo{author}{\bibfnamefont{A.}~\bibnamefont{Zakharov}},
  \bibinfo{author}{\bibfnamefont{A.}~\bibnamefont{Kharitonov}},
  \bibinfo{author}{\bibfnamefont{V.}~\bibnamefont{Shustov}},
  \bibinfo{author}{\bibfnamefont{V.}~\bibnamefont{Kuz'minov}},
  \bibinfo{author}{\bibfnamefont{M.}~\bibnamefont{Lasakov}},
  \bibinfo{author}{\bibfnamefont{R.}~\bibnamefont{Tal'daev}},
  \bibinfo{author}{\bibfnamefont{A.}~\bibnamefont{Aldushchenkov}},
  \bibinfo{author}{\bibfnamefont{V.}~\bibnamefont{Varlamov}},
  \bibnamefont{et~al.}, \bibinfo{journal}{Nuclear Instruments and Methods in
  Physics Research Section A: Accelerators, Spectrometers, Detectors and
  Associated Equipment} \textbf{\bibinfo{volume}{440}}, \bibinfo{pages}{658}
  (\bibinfo{year}{2000}).

\bibitem[{\citenamefont{Atchison et~al.}(2005)\citenamefont{Atchison, Blau,
  van~den Brandt, BryÅ›, Daum, Fierlinger, Hautle, Henneck, Heule, Kirch
  et~al.}}]{atchison2005a}
\bibinfo{author}{\bibfnamefont{F.}~\bibnamefont{Atchison}},
  \bibinfo{author}{\bibfnamefont{B.}~\bibnamefont{Blau}},
  \bibinfo{author}{\bibfnamefont{B.}~\bibnamefont{van~den Brandt}},
  \bibinfo{author}{\bibfnamefont{T.}~\bibnamefont{BryÅ›}},
  \bibinfo{author}{\bibfnamefont{M.}~\bibnamefont{Daum}},
  \bibinfo{author}{\bibfnamefont{P.}~\bibnamefont{Fierlinger}},
  \bibinfo{author}{\bibfnamefont{P.}~\bibnamefont{Hautle}},
  \bibinfo{author}{\bibfnamefont{R.}~\bibnamefont{Henneck}},
  \bibinfo{author}{\bibfnamefont{S.}~\bibnamefont{Heule}},
  \bibinfo{author}{\bibfnamefont{K.}~\bibnamefont{Kirch}},
  \bibnamefont{et~al.}, \bibinfo{journal}{Physical Review Letters}
  \textbf{\bibinfo{volume}{95}}, \bibinfo{pages}{182502}
  (\bibinfo{year}{2005}).

\end{thebibliography}

\end{document}